\documentclass[12pt,preprint]{aastex}
\usepackage{graphicx}

\newcommand\ha{\hbox{H$\alpha$}}
\newcommand\ic{$I_\mathrm{C}$}

\newcommand\ks{$K_\mathrm{s}$}
\newcommand\mum{\hbox{$\mu$m}}
\newcommand\tm{\textit{2MASS\/}}
\newcommand\wise{\textit{WISE\/}}
\newcommand\spitzer{\textit{Spitzer\/}}
\newcommand\co{$^{13}$CO}
\newcommand\sdss{\textit{SDSS\/}}
\newcommand\msun{M$_{\sun}$}

\shorttitle{L1340}
\shortauthors{Kun et al.}

\begin{document}

\title{THE INTERMEDIATE-MASS STAR FORMING REGION LYNDS 1340. \\
 AN OPTICAL VIEW}
\author{M\'aria Kun, Attila Mo\'or, Elza Szegedi-Elek}
\affil{Konkoly Observatory, Research Centre for Astronomy and Earth Sciences, Hungarian Academy of Sciences, H-1121 Budapest, Konkoly Thege \'ut  15--17, Hungary}
\and
\author{Bo Reipurth}
\affil{Institute for Astronomy, University of Hawaii at Manoa, 640 N. Aohoku Place, Hilo, HI 96720, USA}

\date{Received / Accepted}

\begin{abstract}

We have performed an optical spectroscopic and photometric search for young stellar objects associated with the molecular cloud Lynds~1340, and examined the structure of the cloud by constructing an extinction map, based on \sdss\ data. The new extinction map suggests a shallow, strongly fragmented cloud, having a mass of some 3700~M$_{\sun}$. Longslit spectroscopic observations of the brightest stars over the area of L1340 revealed that the most massive star associated with L1340 is a B4 type, $\sim 5$~M$_{\sun}$ star. The new spectroscopic and photometric data of the intermediate mass members led to a revised distance of $825^{+110}_{-80}$\,pc, and revealed seven members of the young stellar population with $M \gtrsim 2$\,\msun. Our search for \ha\ emission line stars, conducted with the {\it Wide Field Grism Spectrograph~$2$\/} on the $2.2$-meter telescope of the University of Hawaii and covering a $30\arcmin \times 40\arcmin$ area, resulted in the detection of 75 candidate low-mass pre-main sequence stars, 58 of which are new. We constructed spectral energy distributions of our target stars, based on \sdss, \tm, \spitzer, and \wise\ photometric data, derived their spectral types, extinctions, and luminosities from {\it BVRIJ\/} fluxes, estimated masses by means of pre-main sequence evolutionary models, and examined the disk properties utilizing the 2--24\,\micron\ interval of the spectral energy distribution. We measured the equivalent width of the \ha\ lines and derived accretion rates. The optically selected sample of pre-main sequence stars has a median effective temperature of 3970\,K, stellar mass 0.7\,M$_{\sun}$, and accretion rate of 7.6$\times10^{-9}$~\msun\,yr$^{-1}$. 
\end{abstract}



\maketitle
                                                                                
\section{INTRODUCTION}
\label{Sect_1}

Lynds~1340 is an isolated dark cloud at ({\it l,b\/}) = (130.1\degr,11.5\degr) \citep{Kun2008}, near the top of the Galactic molecular disk \citep[see e.g. the wide-field 12\,\mum\ \wise\ Sky Survey Atlas (WSSA) image in][]{Meisner14}. According to the first large-scale study of the region \citep[][hereinafter Paper~I]{KOS94} L1340 is located at 600\,pc from the Sun, and the blue reflection nebulosity DG\,9 \citep{DG}, illuminated by a few mid-B and early A-type stars (Paper~I), suggest that it belongs to the class of the intermediate-mass star-forming regions \citep[IM SFRs,][]{Arvidsson2010,Lund14} whose origin, evolution, and role in the Galactic star-forming history are not well studied. IM\,SFRs are not associated with H\,II zones, but illuminate blue reflection nebulae and excite the polycyclic aromatic hydrocarbon molecules of the environment, resulting in strong mid-infrared  diffuse background. The $^{13}$CO observations of L1340 revealed a total mass of 1100\,M$_{\sun}$, some two orders of magnitude smaller than that of a typical giant molecular cloud. Three dense C$^{18}$O clumps, L1340\,A, L1340\,B and L1340\,C comprise some 85\% of the total molecular mass. Three red and nebulous objects, RNO\,7, RNO\,8, and RNO\,9 \citep{RNO}, containing small groups of faint stars, are associated with the three clumps, respectively. Ten dense cores were identified in L1340 through a large-scale NH$_3$ survey by \citet*[][hereinafter Paper~II]{KWT03}, with masses and kinetic temperatures halfway between the values obtained for the ammonia cores in Taurus and Orion \citep{Jijina99}. Thirteen \ha\ emission objects ([KOS94]\,HA\,1--[KOS94]\,HA\,13) were found by \citet{KOS94}, and 14 ones (RNO\,7~HA\,1--14), concentrated in the small nebulous cluster RNO~7 by \citet*{MMN03}. Herbig--Haro objects HH\,487, HH\,488, HH\,489, HH\,671, and their driving sources are reported in \citet*{KAY03} and \citet{MMN03}. 

To explore the nature of interstellar processes, leading to star formation in such an environment, and the role of feedback from young intermediate-mass stars on their natal cloud, the structure of the cloud and the census of the young stellar object (YSO) population have to be assessed. The low sensitivity of the photographic \ha\ survey presented in Paper~I, and the small angular coverage of the more sensitive \ha\ survey by \citet{MMN03} suggest that most of the classical T~Tauri stars associated with L1340 are still undiscovered.
In order to identify the YSO population of L1340 we performed a wide-field slitless grism survey for \ha\ emission stars, and low-resolution, optical longslit spectroscopic observations of the bright stars of the region, illuminating reflection nebulae. To derive the luminosities of the target stars and examine their spectral energy distributions (SEDs) we supplemented our observations with optical photometric data available in the \sdss\ data base, as well as with \textit{2MASS\/}, \textit{Spitzer}, and \textit{WISE\/} infrared photometric data. To study the cloud structure we constructed a new extinction map using \sdss\ data. The results complement our recent infrared search for the YSO population of the cloud, based on our own observations and public data bases \citep[][hereinafter Paper~III]{Kun2016}.  
Our data and analysis are described in Sect.~\ref{Sect_2} and \ref{Sect_3}, respectively. The results are presented in Sect.~\ref{Sect_4}, and discussed in Sect.~\ref{Sect_5}. A brief summary is given in Sect.~\ref{Sect_6}.

\section{DATA}
\label{Sect_2}

\subsection{Longslit Spectroscopy}
\label{Sect_spec}
\begin{figure}
\centerline{\includegraphics[width=16cm]{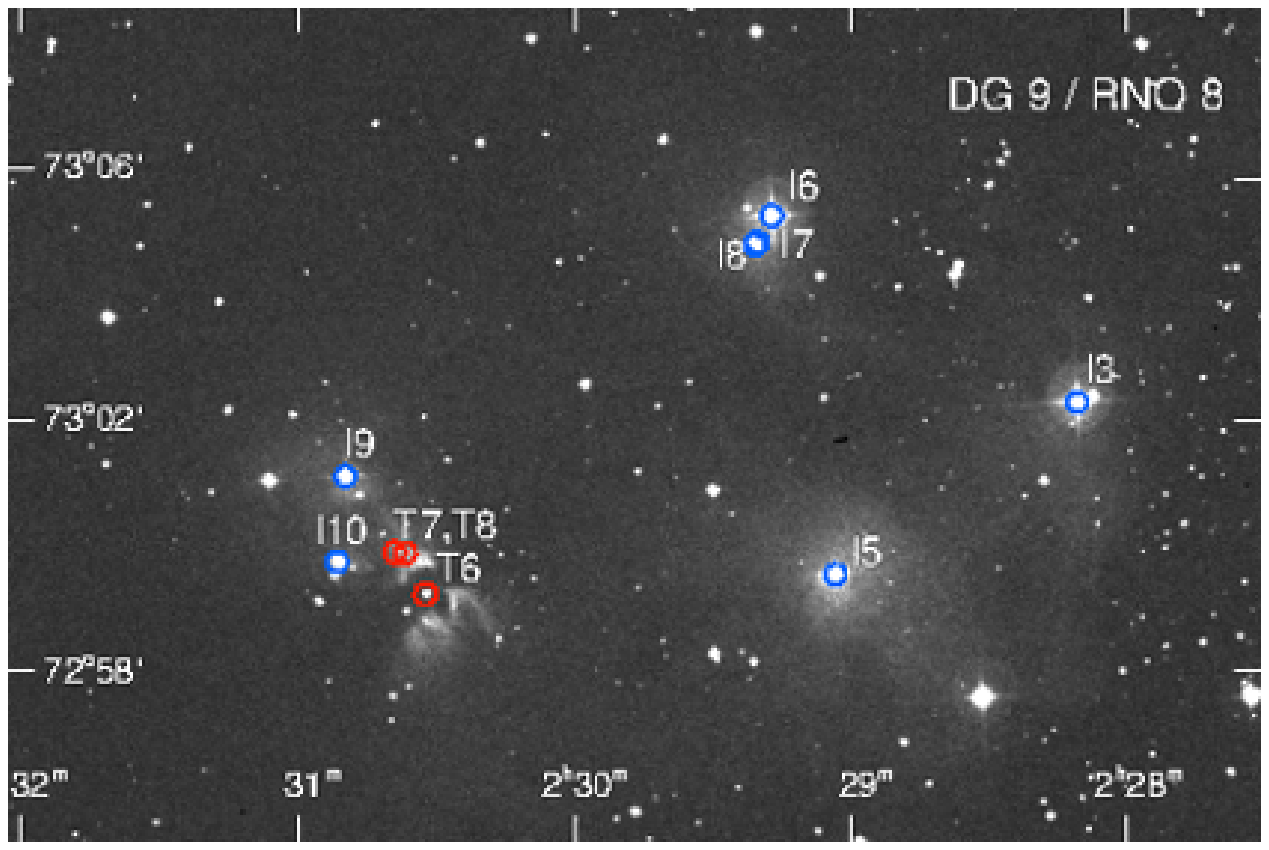}}
\centerline{\includegraphics[width=7.95cm]{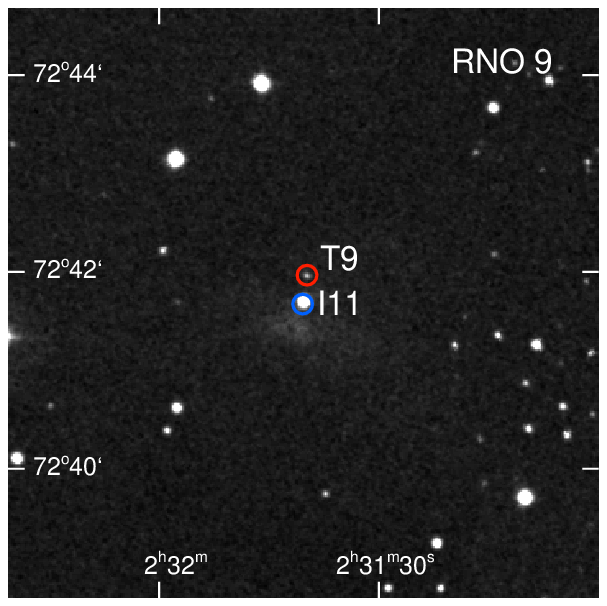}\hskip1mm\includegraphics[width=7.95cm]{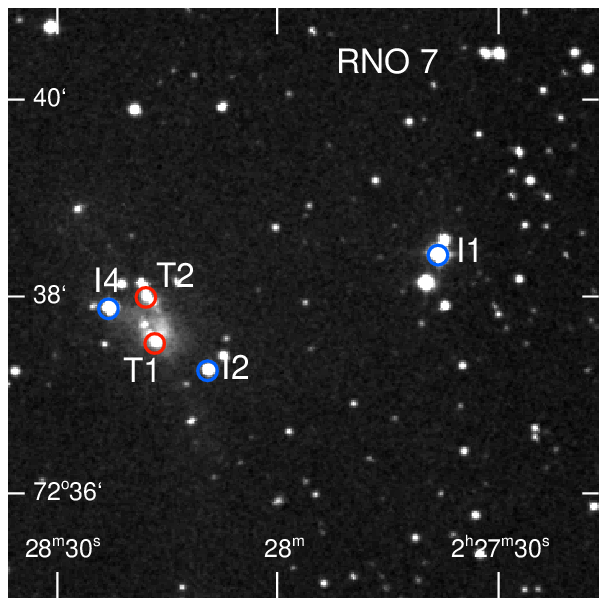}}
\caption{Targets of our spectroscopic observations, marked on the DSS2 blue images. Top: Nebulous stars associated with DG\,9 and RNO\,8; Bottom right: Blue and nebulous stars observed in the RNO\,7 region; Bottom left: the stars in the RNO~9 region. Numbering of the stars are same as in Table~\ref{Table_splog}. Blue and red circles indicate blue and red objects, respectively.}
\label{Fig_sptarget}
\end{figure}

To identify and classify the most luminous optically visible young stars associated with L1340 we obtained low and intermediate resolution optical spectra of seven stars, associated with the extended reflection nebulosity DG\,9 \citep{DG}, three stars in the region of the nebulous cluster RNO\,7, three stars belonging to the small, red nebulosity RNO\,8, and two stars in the RNO\,9 region \citep{RNO}, as well as of six further [KOS94]\,HA stars. Furthermore, we included in the target list two stars projected on the RNO~7 cluster and much brighter and bluer than typical cluster stars. The target stars whose association with the cloud is evidenced by reflection nebulae are shown in Fig.~\ref{Fig_sptarget}. The star No.~1, located east of RNO~7 (lower right panel of Fig.~\ref{Fig_sptarget}, 2MASS~02273807+7238267) was selected due to its associated 8-\mum\ nebulosity (Paper~III).  

We observed the optical spectra of 23 stars utilizing several instruments, namely CAFOS\footnote{\url{http://w3.caha.es/CAHA/Instruments/CAFOS/}} with the G--100 grism, installed on the 2.2-m telescope of the Calar Alto Observatory, FAST on the 1.5-m telescope of the Fred Lawrence Whipple Observatory \citep{Fabricant}, ALFOSC\footnote{\url{http://www.not.iac.es/instruments/alfosc/}} with grism~8 on the Nordic Optical Telescope in the Observatorio del Roque de los Muchachos in La Palma, and the low-resolution slit spectrograph operated on the 1-m RCC telescope of the Konkoly Observatory\footnote{\url{http://www.konkoly.hu/staff/racz/Spectrograph/Medium-resolution.html}}. The log of the spectroscopic observations is shown in Table~\ref{Table_splog}. We reduced the data following the standard IRAF procedures \citep[see further details in][]{Kun2009}. 

\placetable{Table_splog}

\placefigure{Fig_sptarget}

\subsection{Slitless Grism Spectroscopic Observations}
\label{Sect_ha}
We observed L1340 with the Wide Field Grism Spectrograph~2 (\textit{WFGS2\/}), installed on the University of Hawaii 2.2-meter telescope, on 2011 January 1, October 15, 16, and 18, and 2012 August 10. We used a 300\,line\,mm$^{-1}$ grism, blazed at 6500\,\AA, and providing a dispersion of 3.8~\AA~pixel$^{-1}$ and a resolving power of 820. The narrow band \ha\ filter had a 500\,\AA\  passband centered near 6515\,\AA. The detector for WFGS2 is a Tektronix $2048\times2048$ CCD, whose pixel size of 24\,\mum\  corresponds to 0.34~arcsec on the sky. The field of view is $11.5\arcmin\times11.5\arcmin$. We covered an area of $30 \times 40$~arcmin, centered on RA(2000)$= 2^\mathrm{h}30^\mathrm{m}$ and Dec(2000)$=72\degr48\arcmin$, with a mosaic of 12 overlapping fields. For each field, we took a short, 60\,s exposure in order to detect the \ha\ line in bright stars, and three frames of 300~s exposure.

\begin{figure}
\centering{\includegraphics[width=8cm]{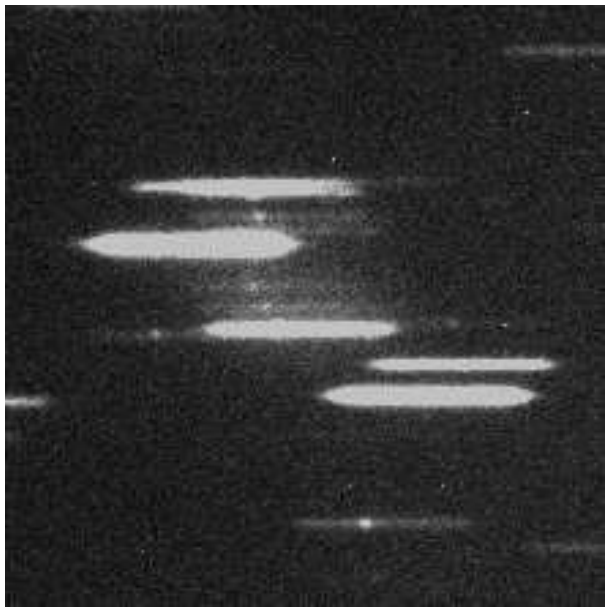}}
\caption{Part of the \textit{WFGS2} image showing the \ha\ emission stars of the RNO~7 cluster.}
\label{Fig_grism}
\end{figure}

Bias subtraction and flat-field correction of the images were done in IRAF. Then we used the FITSH, a software package for astronomical image processing\footnote{http://fitsh.szofi.net/} \citep{pal2012} to remove cosmic rays, coadding the long-exposure images, identify the stars on the images, and transform the pixel coordinates into equatorial coordinate system. As an example of the reduced and coadded images, the central part of the RNO\,7 cluster can be seen in Fig.~\ref{Fig_grism}.

\placefigure{Fig_grism}

\begin{figure*}
\centering{\includegraphics[width=16cm]{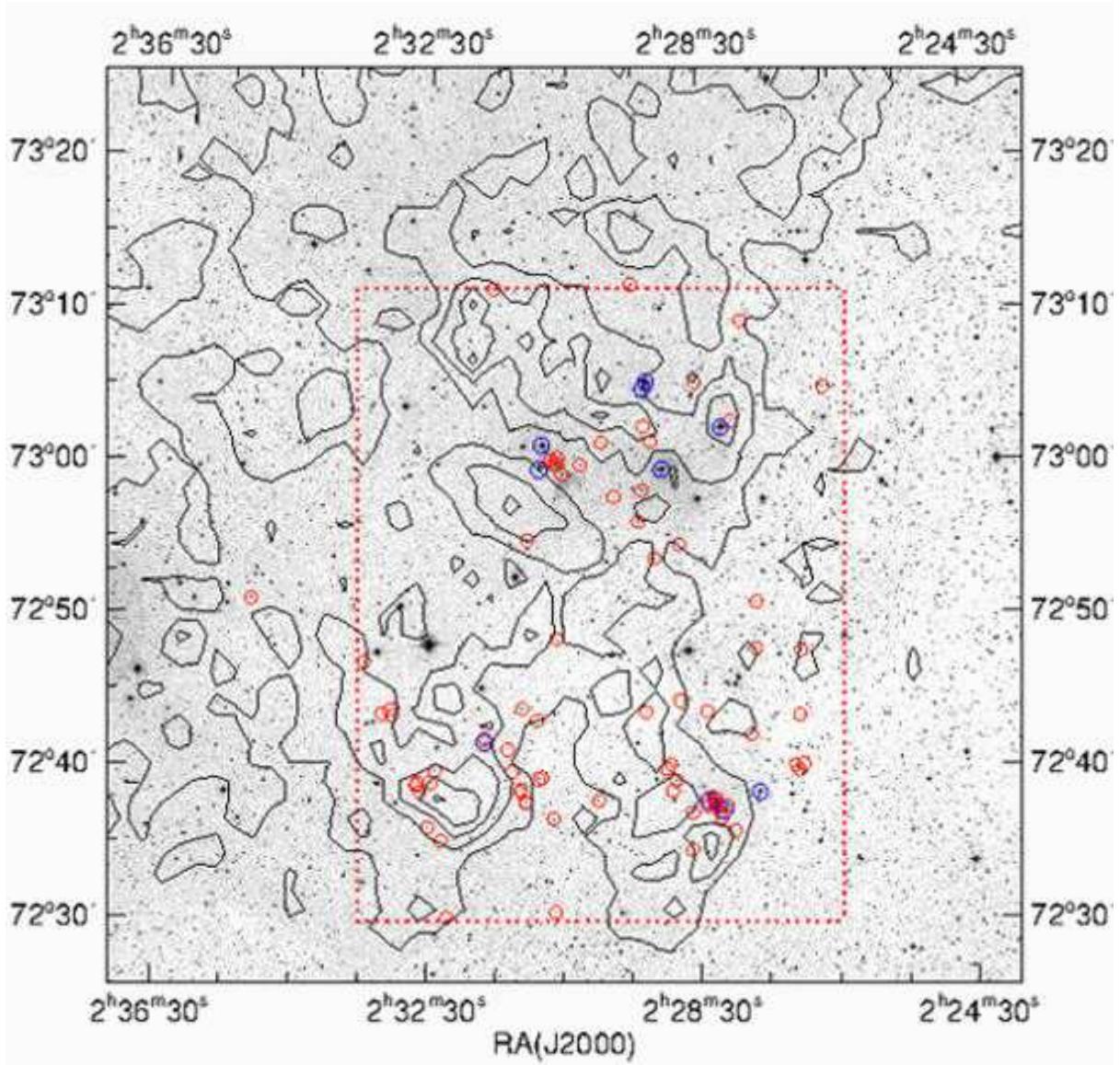}}
\caption{Positions of the \ha\ emission stars (red circles) and stars earlier than F5 (blue circles) on the DSS2 red image of the 1 square degree region centered on RA(J2000.0)=$2^\mathrm{h}30^\mathrm{m}30^\mathrm{s}$, Dec(J2000.0)=$+72\degr56\arcmin$. Extinction contours of $A_\mathrm{V}=1,2,3\ldots$ are overplotted. The dotted rectangle outlines the area covered by the WFGS2 observations.}
\label{Fig_hamap}
\end{figure*}

We found 75 stars with \ha\ emission by examining the images visually. We determined their equatorial coordinates by matching our images with the 2MASS \citep{2mass} image of the field. We could associate all but two emission sources with 2MASS point-sources unambiguously within $1\arcsec$. Therefore we use 2MASS designations of the stars for equatorial coordinates. For the two stars missing from the 2MASS All Sky Catalogue we use their SDSS DR9 identifiers. One of these stars, SDSS9~J022856.42+724019.2, situated at some 2\arcsec\ from another, brighter \ha\ emission star, 2MASS~02285635+7240171 (SDSS9~J022856.34+724017.1), has no counterpart in any of the \tm, \wise, and \spitzer\ data. The other one, SDSS9~J022932.32+725503.3, although too faint for the \tm, can be identified in the AllWISE data base and \spitzer\ images.
We detected \ha\ emission from 18 stars, previously identified as \ha\ emission objects by \citet{KOS94} and \citet{MMN03}, and from the hypothetical exciting source of the Herbig--Haro object HH\,488 \citep[HH\,488\,S,][]{KAY03}. The equivalent width of the H$\alpha$ emission line EW(\ha) and its uncertainty were computed in the manner described by \citet{Szeg13}. Due to the faint continuum or overlapping spectra we could not measure EW(\ha) in the spectra of six stars. Table~\ref{Table_halpha1} lists the \tm\ designations, measured EW(\ha), and cross-identifiers of the \ha\ emission stars of L1340, consisting of the 75 stars identified in the WFGS2 images and the two ones identified during the longslit observations but outside of the field of view of the WFGS2 observations.  Figure~\ref{Fig_hamap} shows the positions of the \ha\ emission stars, overplotted on the DSS2 red image of the region.

\placetable{Table_halpha1}

\placefigure{Fig_hamap}

\subsection{Photometric Data}
\label{Sect_phot}
\paragraph{\textit{Spitzer}} L1340 was observed by the \textit{Spitzer Space Telescope} using \spitzer's Infrared Array Camera  \citep[IRAC;][]{Fazio2004} on 2009 March 16 and by the Multiband Imaging Photometer for Spitzer   \citep[MIPS;][]{Rieke2004} on 2008 November 26 (Prog. ID: 50691, PI: G. Fazio). The IRAC observations covered $\sim 1$~deg$^2$ in all four bands. The centers of the 3.6 and 5.8~\mum\ images are slightly displaced from those of the 4.5 and 8~\mum\ images, therefore part of the clump L1340\,C is outside of the 4.5 and 8~\mum\ images. Moreover, the 24 and 70\,\mum\ images do not cover the southern half of L1340\,A. A small part of the cloud, centered on RNO~7, was observed in the four IRAC bands on 2006 September 24 (Prog. ID: 30734, PI: D. Figer). Each of our longslit spectroscopic target stars  are located within the field of view of the \spitzer\ observations. All but two of the 75 \ha\ emission stars detected by the WFGS2 are within the field of view of the 3.6 and 5.8-\mum\  images, albeit 14 of them are outside of the 4.5 and 8-\mum\ images. We performed IRAC and MIPS photometry of the target stars by the procedure described in \citet{Kun2014}. 

\paragraph{\textit{AllWISE\/}}  All but eight of the \ha\ sources have counterparts in the \textit{AllWISE\/} Source Catalog \citep{Wright2010}. The \wise\  images of L1340 reveal bright diffuse background at 12 and 22~\micron\ (see Paper~III). Point source fluxes in the \textit{W3} and \textit{W4} bands may therefore be contaminated by the diffuse radiation, originating from the environment. We applied the criteria, set by \citet{KL2014}, to discriminate real point source fluxes and fake sources, resulting from background-contaminated data in the \textit{W3} and \textit{W4} bands. We found reliable \wise\ 22-\mum\ fluxes for five \ha\ emission stars, not observed or not detected by MIPS at 24-\mum, allowing us to classify the SEDs of these stars. 

\paragraph{\textit{SDSS\/}} L1340 is situated within Stripe~1260 of the {\it SEGUE\/} survey \citep{Yanny2009}. Each of our target stars has a counterpart in the SDSS Data Release~9 \citep{Ahn2012} within 1\arcsec\ of the 2MASS position. We include the optical data points into the SEDs of the stars, and use them for estimating their extinctions and spectral types. To compare the color indices with those of the spectral sequence of pre-main sequence stars, published by \citet{Pecaut2013}, we transformed the \textit{ugriz} magnitudes into the Johnson--Cousins {\it UBVR}$_\mathrm{C}${\it I}$_\mathrm{C}$  system, using the equations given in \citet[][for {\it BVR}$_\mathrm{C}${\it I}$_\mathrm{C}$]{Ivezic}  and \citet[][for {\it U}]{Jordi06}. Furthermore, we use \textit{SDSS\/} data for constructing an extinction map of the cloud. 

Photometric data of the target stars are presented in Tables~A1 and A2 of Appendix~A. Table~A1 lists the {\it UBVR}$_\mathrm{C}${\it I}$_\mathrm{C}$ and \textit{2MASS\/} \textit{JHK$_s$} magnitudes,  and Table~A2  contains the \textit{Spitzer\/} [3.6], [4.5], [5.8], [8.0], [24], and [70.0] magnitudes, and \textit{AllWISE\/} [3.4], [4.6], [12], and [22] magnitudes for the optically selected candidate young stars associated with L1340. 

\section{ANALYSIS}
\label{Sect_3}

\subsection{Spectral Classification}
\label{Sect_spclass}
\begin{figure*}
\centering{
\includegraphics[width=16cm]{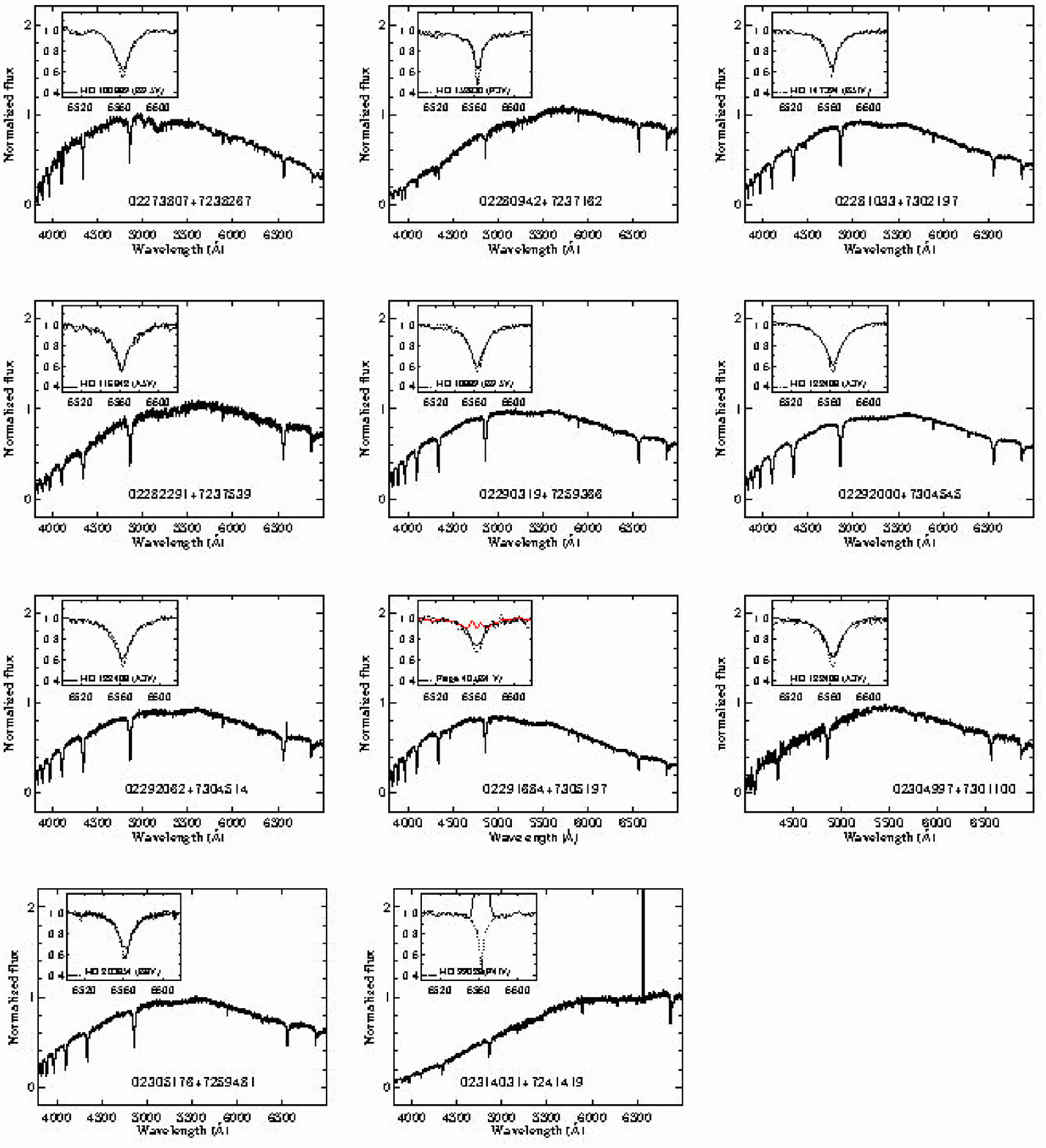}}
\caption{Spectra of the observed stars earlier than F5. In the inset the region of the \ha\ line is magnified. The red line in the inset of the star 02291684+7305197  shows the \ha\ line observed on 1999 August 7, and the black line shows the spectrum detected on 2004 November 6. For comparison the \ha\ line of a standard star of the same type is drawn with a dotted line.}
\label{Fig_splotab}
\end{figure*}

\begin{figure*}
\centering{
\includegraphics[width=16cm]{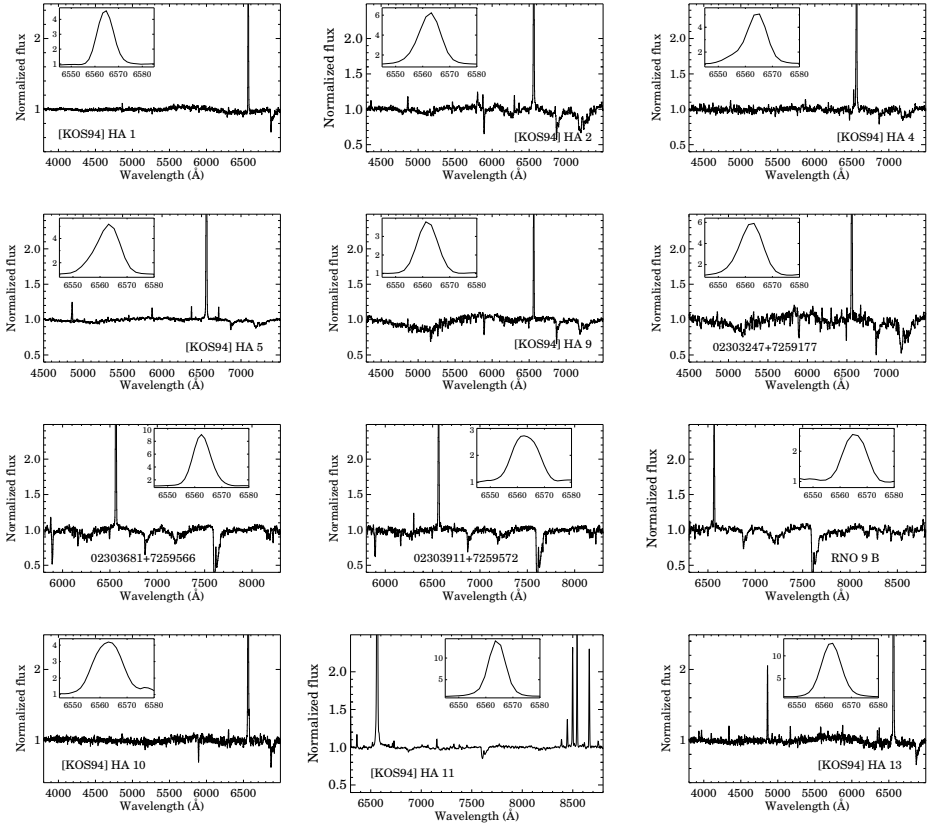}}
\caption{Spectra of [KOS94]\,HA stars and nebulous stars associated with RNO~7, RNO~8 and RNO~9. Insets show a zoom on the \ha\ lines.}
\label{Fig_splotha}
\end{figure*}

We analysed our longslit spectra using the `splot' task of IRAF. The spectral types of the stars were determined by comparing the absorption features with those in a number of standard star spectra found in the spectrum library of \citet*{JHC84}, and following the criteria described in \citet{Hernan} and \citet{Gray}. Spectra of the stars earlier than F5 are shown in Fig.~\ref{Fig_splotab}, and their spectral types are listed in Table~\ref{Table_abf}. We examined the \ha\ lines in these spectra for the presence of a possible emission component, overlying the photospheric \ha\ absorption line. The inset in each panel shows the \ha\ line of the target star together with that of a standard star of the same spectral type. We estimate the accuracy of the spectral types of these stars as $\pm1$ subclass. Spectra of the observed \ha\ emission stars are displayed in Fig.~\ref{Fig_splotha}. The classification of these stars is less accurate due to their low brightness and non-photospheric line and continuum emission. The spectral types of these stars are listed in Table~\ref{Table_spres}. We adopt a two-subclass uncertainty, which is supported by the comparison of the results with those obtained from the SEDs (see Sect.~\ref{Sect_SED}).  

\placetable{Table_abf}
\placetable{Table_spres}

\placefigure{Fig_splotab}
\placefigure{Fig_splotha}

\subsection{Spectral Energy Distributions}
\label{Sect_SED}
The available \sdss, \tm, \spitzer, and \wise\ photometric data allowed us to plot the SEDs of the observed stars over the 0.36--24\,(70)\,\micron\ wavelength region. 
Having the spectral types of the longslit target stars determined, we dereddened their observed SEDs according to the normal interstellar reddening law ($R_\mathrm{V} =3.1$) of \citet*{CCM89} to match the photospheric SED of the given spectral type, defined by the color indices tabulated by \citet{Pecaut2013}. 

SEDs of the \ha\ emission stars, detected by WFGS2, were also constructed using all photometric data. \textit{Akari\/} IRC fluxes \citep{Ishihara} at 9.0 and 18.0~\mum\ are also included when available. We estimated the spectral type and the extinction of each target star by comparing visually the optical--near infrared SED (from the {\it B\/} to the {\it J\/} band) with those of a grid of reddened photospheres, using the reddening-free color indices of \citet{Pecaut2013}, the extinction law of \citet{CCM89}, and the $A_\mathrm{V} \ge 0.5$\,mag restriction \citep{KWT03}. We found the best fit of the photometry and photospheric colors using the reddening law $R_\mathrm{V}=3.1$ for $A_\mathrm{V} \lesssim 2.0$, and with $R_\mathrm{V}=5.5$ for $A_\mathrm{V} \ga 2.0$. Similar dependence of the extinction law on the line-of-sight $A_\mathrm{V}$ was reported by \citet{Allen2014} for the Cepheus~OB3 star-forming region. We estimate the accuracy of the resulting spectral types and extinction as $\pm2$ subclass and $\pm0.5$~mag, respectively.

We classified the infrared excesses of the \ha\ emission stars based on their dereddened SED slopes, $\alpha=d\log(\lambda F(\lambda)) / d\log\lambda$, over the 2--24\,(22)\,\mum\ interval  (\textit{WISE\/} data were used for classification when no \textit{Spitzer} data were available). According to the canonical classification scheme \citep{Lada,Greene} Class~I protostars are characterized by $\alpha > 0.3$, $-0.3 \le \alpha \le 0.3$ indicates Flat~SED sources, near the boundary between the protostellar and pre-main sequence evolutionary phases, whereas classical T~Tauri stars have Class~II slopes with $-0.3 > \alpha \ge -1.6$. Class~III young stars with $\alpha \le -1.6$ are pre-main sequence stars with very weak or no infrared excess.  Class~II SEDs then can be further divided into the primordial (II\,P), pre-transitional/transitional (II\,T), and weak or `anemic' (II\,A) subclasses \citep{Evans09}, based on the details of the SED over the 2--24\,\micron\  wavelength interval, compared to those of the median SED of the benchmark sample of T~Tauri stars of the Taurus region.  We constructed the Taurus median SED using  \citeauthor{Furlan06}'s \citeyearpar{Furlan06} data, established for K5--M2 type stars over the 1.25\,$\micron\ \le \lambda \le 34.00$\,\micron\  region, and those of \citet{DAlessio} for optical and far-infrared wavelengths. A circumstellar disk is primordial if the SED does not drop below the Taurus median band; it is an evolved pre-transitional or transitional disk if the SED is below the Taurus median band in the near-infrared, and starts rising in the mid-infrared, whereas the SED of a weak or anemic disk is below the Taurus band over the whole observed infrared region. Since the evolutionary processes leading to the II\,A and II\,T  subclasses may be different, their occurrence and properties within the same star-forming environment may bear information on their origin.  

\subsection{Hertzsprung--Russell Diagram}
\label{Sect_hrd}
Having determined spectral classes and extinctions we calculated the bolometric luminosities of the target stars from the extinction-corrected {\it V\/}, \ic, and {\it J\/} magnitudes separately, using the bolometric corrections and color indices tabulated by \citet{Pecaut2013}, and adopting the new distance of 825\,pc (see Sect.~\ref{Sect_825}). Then we took the average of the luminosities obtained from the three photometric bands. To find the positions of our target stars in the $\log T_\mathrm{eff}$\,--\,$\log L$ plane, effective temperatures, corresponding to their spectral types, were also adopted from \citet{Pecaut2013}. 
To estimate the masses and ages of our target stars, we applied evolutionary tracks and isochrones from \citeauthor{siess}'s \citeyearpar{siess} pre-main sequence evolutionary models. The errors of $T_\mathrm{eff}$ originate from the uncertainty of the spectral classification, and the errors of the luminosities were propagated from the photometric errors, and uncertainties of distance, extinction and bolometric correction. 

\subsection{Extinction Mapping}
\label{Sect_extmap}
\begin{figure*}
\centering{\includegraphics[width=8cm]{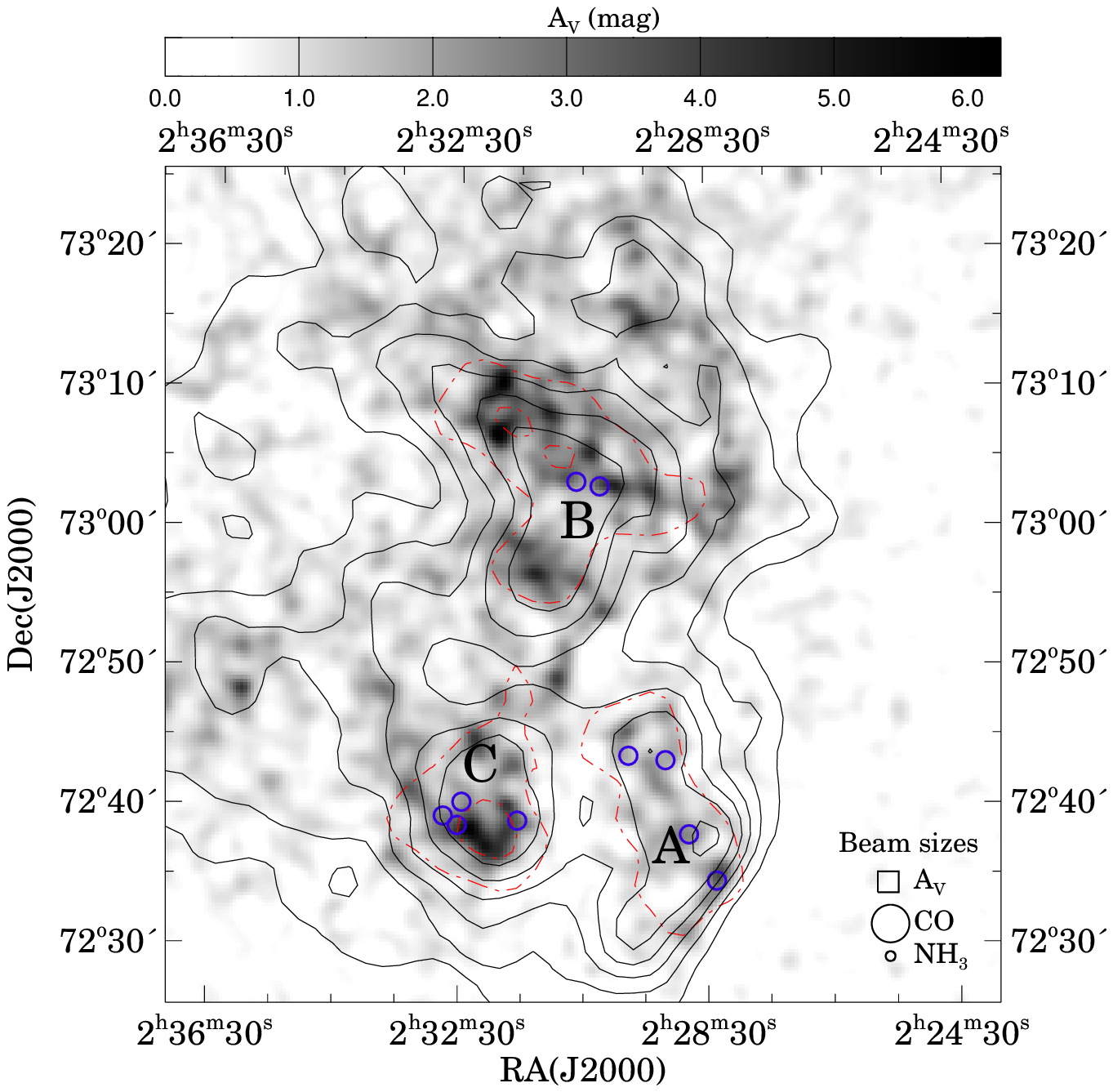}\hskip1cm
\includegraphics[width=8cm]{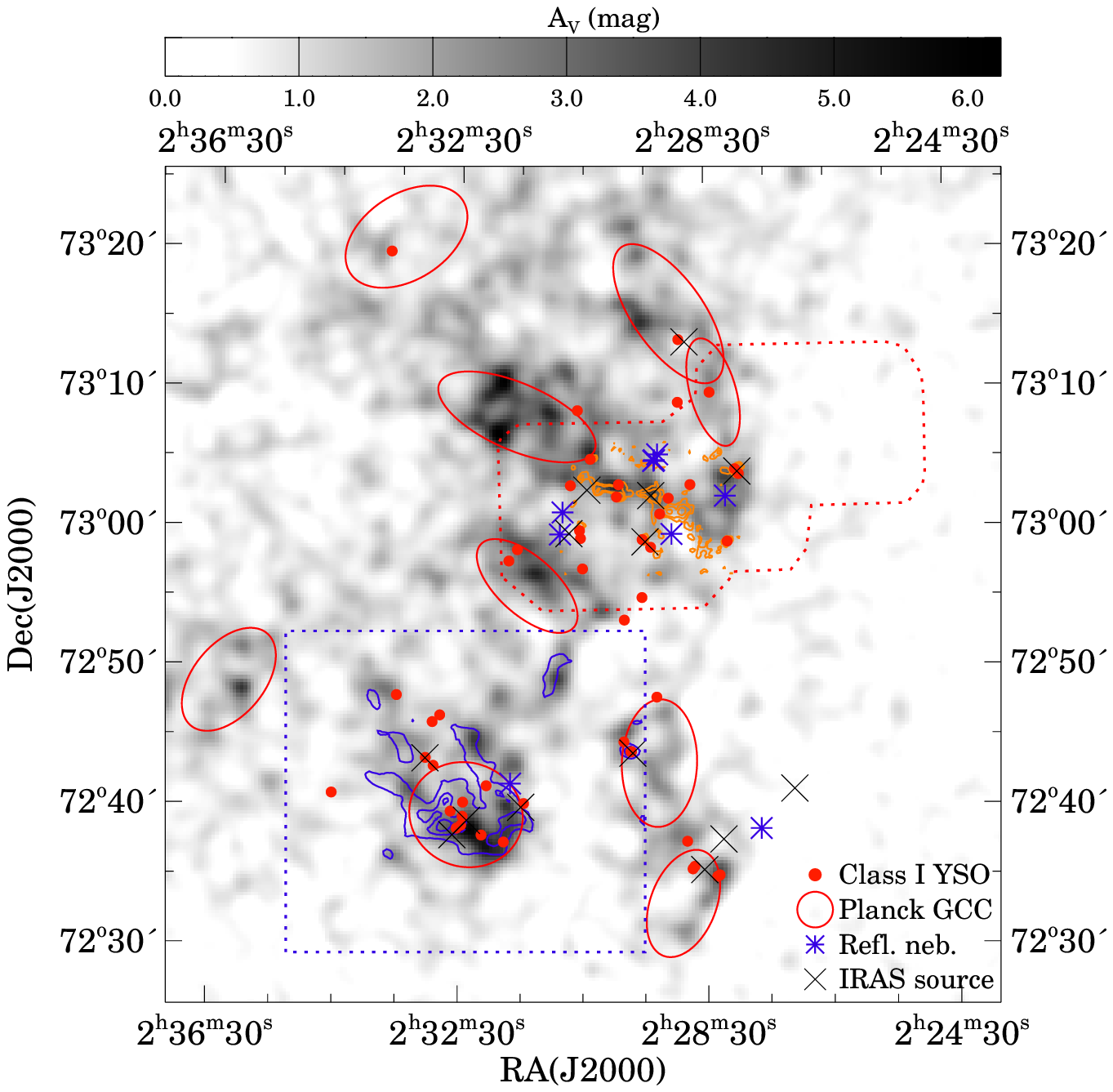}}
\caption{Visual extinction map of L1340, determined from star counts in the \textit{SDSS} DR9. (a) Thin solid black contours show the integrated intensity contours of the \co. The lowest contour and the increment are 0.56\,K\,km\,s$^{-1}$. The red dash-dotted lines indicate the C$^{18}$O contours $T_A=0.30 (\sim 5\,\sigma)$, and 0.60\,K. Blue circles show the positions of NH$_3$ cores (Paper~II). (b) Comparison of the extinction map with various dust-tracers. The lowest contour is at $A_\mathrm{V}=1$\,mag, and the increment is 1\,mag. Yellow contours indicate 850\,\mum\ continuum detected by \textit{SCUBA}. Contours are at 25, 50,, and 100 mJy\,beam$^{-1}$. Blue contours show the 500-\mum\ emission of the cold dust, measured by the SPIRE instrument of the \textit{Herschel} Space Observatory \citep{Juvela}. The levels are at 30, 50, and 70 Jy\,pixel$^{-1}$. Ellipses indicate the positions and sizes of the \textit{Planck\/} cold cores. Reflection nebulae and embedded YSOs are also overplotted. Meaning of the symbols are shown in the lower right corner. The red dotted line indicates the approximate boundary of the \textit{SCUBA} observations, and the blue dotted line borders the \textit{Herschel} 500\,\mum\ image.}
\label{Fig_extmap}
\end{figure*}

The high sensitivity \sdss\ data and the improved census of the YSO population of L1340 allow us to refine the picture of the dust column density structure of L1340, compared to available extinction maps of the region \citep[Paper~II,][]{RF2009,TGU}. We constructed an extinction map, applying the classical method of star counts \citep{Dickman} on the SDSS~DR9 data set. We repeated the procedure that was applied on the DSS2 data in Paper~II. We counted the stars on 90-arcsec sized squares, whose centres were distributed on a regular grid with step of 15\arcsec. The number of stars with {\it V} $\le 25$~mag within a 1 square degree area was 30562. We omitted all classified galaxies, and removed each star with {\it V} $< 14$~mag as probable foreground object. Furthermore, we removed each identified candidate YSO, both the \ha\ emission stars and the color-selected pre-main sequence stars (Paper~III). The off-cloud reference area was a $6\arcmin\times6\arcmin$ field centered at $RA(2000)=34.4779\degr, D(2000)=+72.993\degr$, containing 499 stars. The resulting $A_\mathrm{V}$ map then was boxcar-smoothed to reduce the scatter. The uncertainty of the extinction of a pixel, derived by the formula given in \citet{Dickman} and depending on the pixel value itself, extends from $\sim 0.4$~mag in the low extinction areas (between 0--1\,mag) to 1.4\,mag near the extinction peaks. The resulting $A_\mathrm{V}$ map can be seen in Fig.~\ref{Fig_extmap}.

\placefigure{Fig_extmap}

\section{RESULTS} 
\label{Sect_4}

\subsection{Revised Distance}
\label{Sect_825}
Main-sequence stars illuminating reflection nebulae are excellent distance indicators of interstellar clouds. The distance of L1340, given in Paper I, was determined using objective-prism spectral classes and photoelectric \textit{UBV} magnitudes of three stars of L1340\,B, associated with reflection nebulae (stars 3, 5, and 6 in Table~\ref{Table_splog}, denoted in Paper~I as R1, R2, and R3, respectively). This method resulted in $660\pm30$\,pc, and it was averaged with the lower value of 560\,pc, suggested by a Wolf diagram, to 600\,pc. We note that, in addition to the limited precision of spectral classification from very low dispersion spectra, the photoelectric magnitudes of some measured stars were contaminated by the light from visual companions, biasing the result toward a smaller distance. The new spectroscopic and photometric data allow us to refine the distance determination of L1340. 

The results of our spectral classification, listed in Table~\ref{Table_abf}, show that the two most luminuous stars associated with L1340 are the B5 type star No.~I3  and the B4 type No.~I6. Both stars were classified as B5 type in Paper~I. The new spectra show that the \ion{He}{1} lines are slightly stronger and the \ion{He}{1}\,$\lambda\,4471$ to \ion{Mg}{1}\,$\lambda\,4481$ ratio is higher in the spectrum of the latter star, suggesting an earlier type. 
Due to the short evolutionary time scales of such stars both of them were most probably born in L1340, and are located on the zero-age main sequence (ZAMS) of the Hertzsprung--Russell diagram. The effective temperatures of these spectral types, according to \citet{Pecaut2013}, are 15700\,K and 16700\,K, respectively, and the luminosities on the ZAMS, according to the intermediate-mass pre-main sequence evolutionary models of \citet{PS93}, are 458\,$L_{\sun}$ and 670\,$L_{\sun}$, respectively. Applying the unreddened color indices and bolometric corrections, tabulated by  \citet{Pecaut2013}, we obtain distances of 855\,pc and 794\,pc for the two stars, respectively. We adopt the new distance of 825\,pc in the rest of this paper. A $\pm1$~subclass scatter of the spectral types results in the distance range $750 \la D \la 940$\,pc. The recent paper of \citet{Green2015} suggests similar distance of the dark clouds at the Galactic coordinates of L1340. 

\subsection{Cloud Structure and Clump Masses}
\label{Sect_cloud}
 The new $A_\mathrm{V}$ map of L1340 is shown in Fig.~\ref{Fig_extmap}. To compare the distribution of the gas and dust, \co\  and C$^{18}$O contours (Paper~I) are drawn and positions of the ammonia cores are overplotted in the left panel (a). The right panel (b) compares the new extinction map with the distribution of various  dust-indicators. We have drawn the contours of the 850-\mum\ emission, available for the clump L1340\,B in the \textit{SCUBA Legacy Catalogues} \citep{DiFran}, and 500-\mum\ emission contours, measured by \textit{Herschel\/} SPIRE for L1340\,C \citep{Juvela}, as well as positions of stars illuminating reflection nebulae (Paper~I), \textit{Planck} Galactic cold cores \citep[PGCCs,][]{Planck2015}, and embedded protostars (Paper~I, Paper~III). 

Our extinction map saturates at $A_\mathrm{V,max} = 6.7$~mag. The extinction exceeds this value on 2 pixels within the area of clump ~A, on 50 pixels within clump~B, and 27 pixels in clump~C. The column density at these positions is close to the critical lower limit of $A_\mathrm{V} \approx 8$~mag found for star-forming clumps in several molecular clouds \citep[e.g.][and references therein]{Molinari14}.  

The overall structure and the mean $A_\mathrm{V}$ are compatible with those obtained by \citet{RF2009} and \citet{TGU}, based on \tm\ color excesses. We find the mean extinction $\langle A_\mathrm{V} \rangle = 1.87$~mag for the region of the cloud within the  $A_\mathrm{V} = 1.0$~mag contour. The same average is 1.85~mag for \citeauthor{RF2009}'s \citeyearpar{RF2009} extinction map, and 1.63 for the map of \citet{TGU} \citep[for the offset of both all-sky extinction maps see][]{RF2009}. A conspicuous difference is that, whereas the most opaque spots of our map are found in the largest clump B, the color index-based method indicates the maximum extinction at the position of the cluster RNO~7. The large median \tm\  color indices at the position of RNO~7, leading to the high apparent extinction, may be caused by the large number of unidentified YSOs (Paper~III). Another difference is that the  maximum $A_\mathrm{V}$, derived from the color excesses, is 6.43~mag, whereas a few small, even darker regions can be detected in the map based on \sdss\ data.  

Comparison of \co\ and $A_\mathrm{V}$ shows that the three clumps L1340\,A, B, and C, revealed by the \co\ data, fragment into dark knots with typical sizes of a few arcminutes in the extinction map. In particular, the central region of L1340\,B contains a conspicuous chain of dark spots, comparable in angular size with the resolution of the map, and also apparent in the 850\,\mum\ emission. 

We estimated the mass of the cloud and its clumps using the extinction map. Assuming $N(HI+2H_{2}) = 1.9\times10^{21}\times A_\mathrm{V}$, 36\% helium and 1\% dust mass, we obtain $M \approx 3700$\,M$_{\sun}$ for the cloud material above $A_\mathrm{V}=1.0$~mag, corresponding to the column density $N(HI+2H_{2}) \ga 0.94\times10^{21}$\,cm$^{-2}$. This part of the cloud covers an area of some 90\,pc$^{2}$, and roughly coincides with the region detected in \co. Scaling the mass of 1100\,\msun, obtained from the \co\ observations, to 825\,pc we obtain 2080\,\msun, demonstrating that the \co\ emission saturates at lower column density than the $A_\mathrm{V}$ map.

We estimated the masses of the clumps by summarizing the column densities for the regions above $A_\mathrm{V} = 1.8$\,mag, where the clumps separate. Table~\ref{Table_clump} lists the average sizes (square root of the area above the threshold), mean and maximum $A_\mathrm{V}$, and masses of the three clumps, adopting the distance of 825\,pc.

\placetable{Table_clump}

Nine \textit{Planck\/} Galactic cold cores are projected near extinction peaks, suggesting their physical relation to L1340. One of them, PGCC~G130.38+11.26 was included in the detailed \textit{Herschel\/} study by \citet{Juvela}. Assuming a distance of 810~pc they derived a mass of 404\,\msun\  for a region of 0.94~pc (4\arcmin) in diameter. Comparison of this result with those in Table~\ref{Table_clump} supports that the mass derived from the dust emission is compatible with that obtained from the extinction. 

\subsection{Intermediate-mass Young Stars}
\label{Sect_iMM}

\subsubsection{Stars Earlier than F5}
\label{Sect_abf}
\begin{figure*}
\centering{\includegraphics[width=16cm]{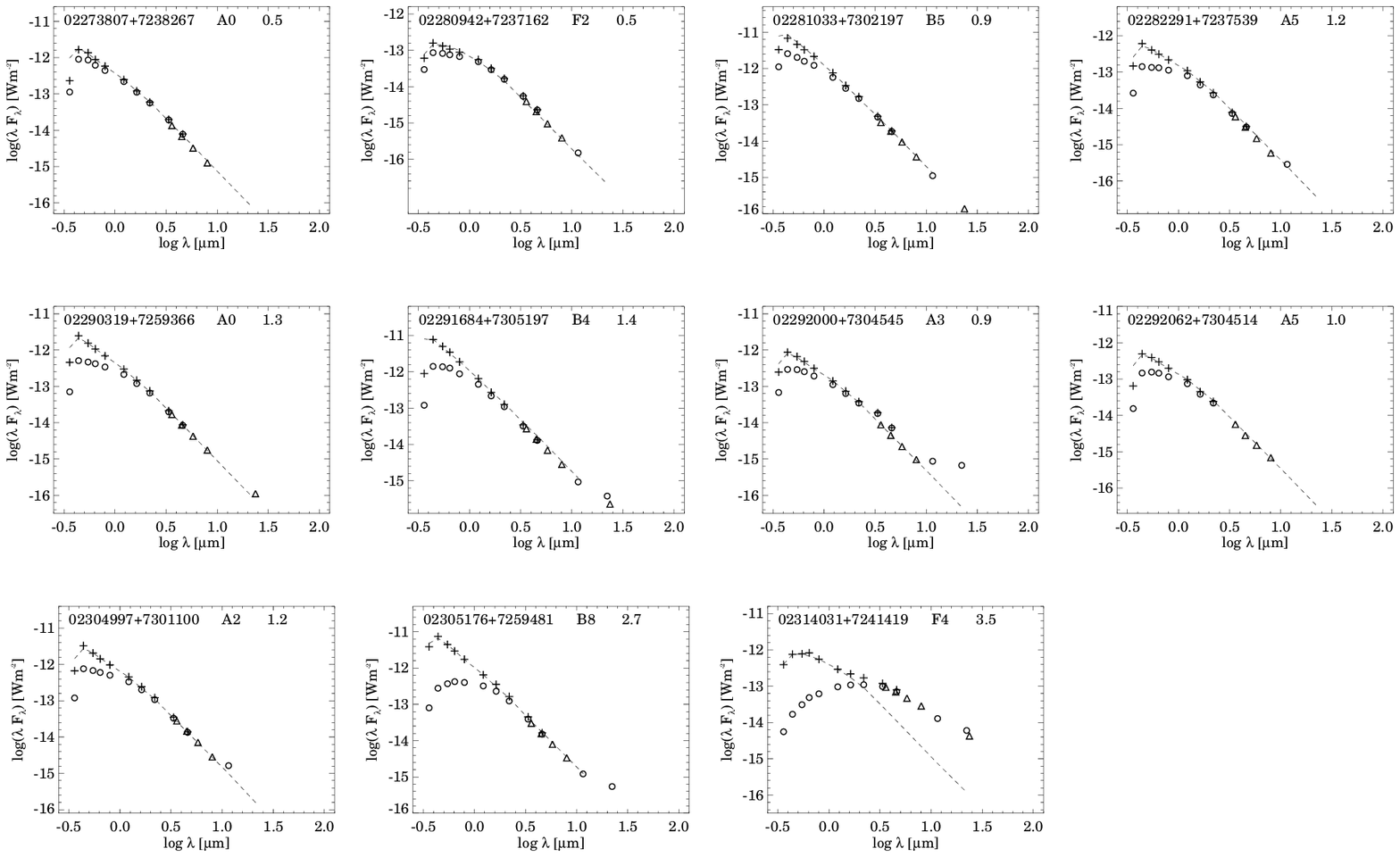}}
\caption{SEDs of the stars earlier than F5 and associated with optical or infrared nebulosities. Open circles show the \textit{SDSS}, \textit{2MASS}, and \textit{WISE} data, and triangles result from our \spitzer\ photometry.  Plusses indicate the dereddened SED, and the dashed line shows the photospheric SED of the spectral type, derived from our spectroscopic data, matched to the extinction-corrected \ic\ band flux. \textit{2MASS} identifiers, spectral types and derived $A_\mathrm{V}$ extinctions are given at the top of each panel.}
\label{Fig_sed_abf}
\end{figure*}

We identified 11 stars earlier than F5, projected on the surface of the molecular cloud. We expect to find optically visible intermediate mass ($M \ga 2$\,$M_{\sun}$) members of the YSO population of L1340 among these stars. The SEDs of these 11 stars, including all available photometric data, are shown in Fig.~\ref{Fig_sed_abf}. The photospheric SEDs, matched to the extinction-corrected data at the \ic\  and \textit{J\/} bands are also plotted. 
Spectral types, extinctions, effective temperatures, and luminosities of these stars are listed in Table~\ref{Table_abf}, and they are plotted with blue star symbols in the Hertzsprung--Russell diagram in Fig.~\ref{Fig_hrd}. 

Five stars, Nos.~I1, I5, I9, I10, and I11 in Table~\ref{Table_abf},  are located above the ZAMS. These stars appear to be 1--3 million year old stars of 2--3~\msun\ mass. Figures~\ref{Fig_splotab} and \ref{Fig_sed_abf} suggest that most of these stars lack emission lines and infrared excesses, characteristic of Herbig~Ae/Be stars. The only obvious exception is star No.~I11, an F4-type star associated with RNO\,9, exhibiting both infrared excess and strong \ha\ emission. Moreover, a weak emission component can be seen in the \ha\ line of the B8-type star No.~10, and the B4-type star No.~I6 exhibited \ha\ emission in the spectrum recorded on 1999 Aug 7 (drawn by red line in Fig.~\ref{Fig_splotab}). The mass of this latter star, estimated from the position on the zero-age main sequence, is about 5~\msun. The two non-nebulous target stars projected on RNO~7 (Nos. I2 and  I4) are located on the ZAMS at the distance of the cloud, like Nos. I7 and I8, members of a visual double embedded in the DG\,9 nebula. 

\placefigure{Fig_sed_abf}

\subsubsection{Intermediate-mass T Tauri Stars}
\label{Sect_imtts}
Figure~\ref{Fig_hrd} suggests that seven of the brightest \ha\ emission stars, (Nos. T1, T4, T8, T10, and T12 in Table~\ref{Table_spres}, as well as Nos. 19 and 20 in Table~\ref{Table_halpha2}) are as massive as 2--2.5~\msun. According to the overplotted evolutionary tracks these stars are in a short-lived phase of their pre-main sequence evolution, and evolving toward higher temperatures and luminosities they will become either Herbig~Ae stars or normal late B--early A-type stars \citep{Herbig94,Calvet2004}.

A candidate embedded intermediate-mass young star is the central star of the optical nebulosity RNO~8, associated with IRAS~02259+7246 (No.  T6 in Table~\ref{Table_spres} and No.~44 in Table~\ref{Table_halpha2}). Its position near the ZAMS suggests that, at optical wavelengths we observe its photospheric spectrum scattered from the outer disk atmosphere, being thereby strongly attenuated \citep{DeMarchi2013}. If we compute the bolometric luminosity of this source by integrating the whole available SED, and assume that the infrared fluxes originate from reprocessed starlight, the star will move upwards in the HRD, to the position indicated by an arrow in Fig.~\ref{Fig_hrd}, suggesting a young star of some 2--2.5~M$_{\sun}$.

\subsection{The Classical T~Tauri Population}
\label{Sect_halpha}
\begin{figure}
\centering{\includegraphics[width=6cm]{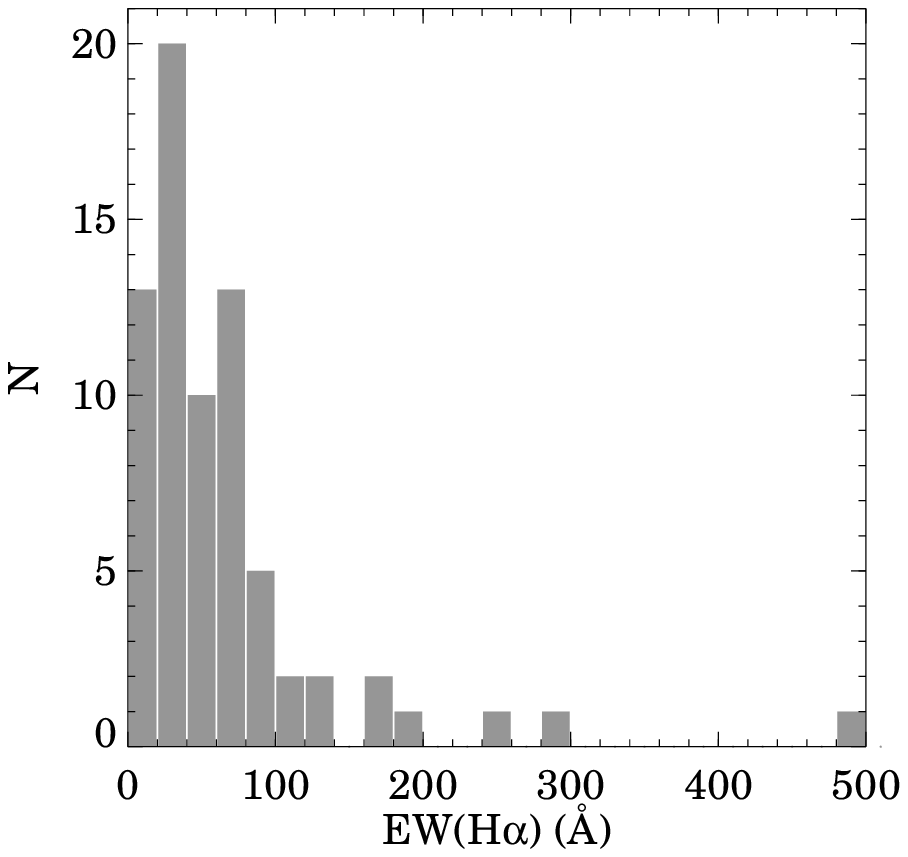}
\includegraphics[width=6cm]{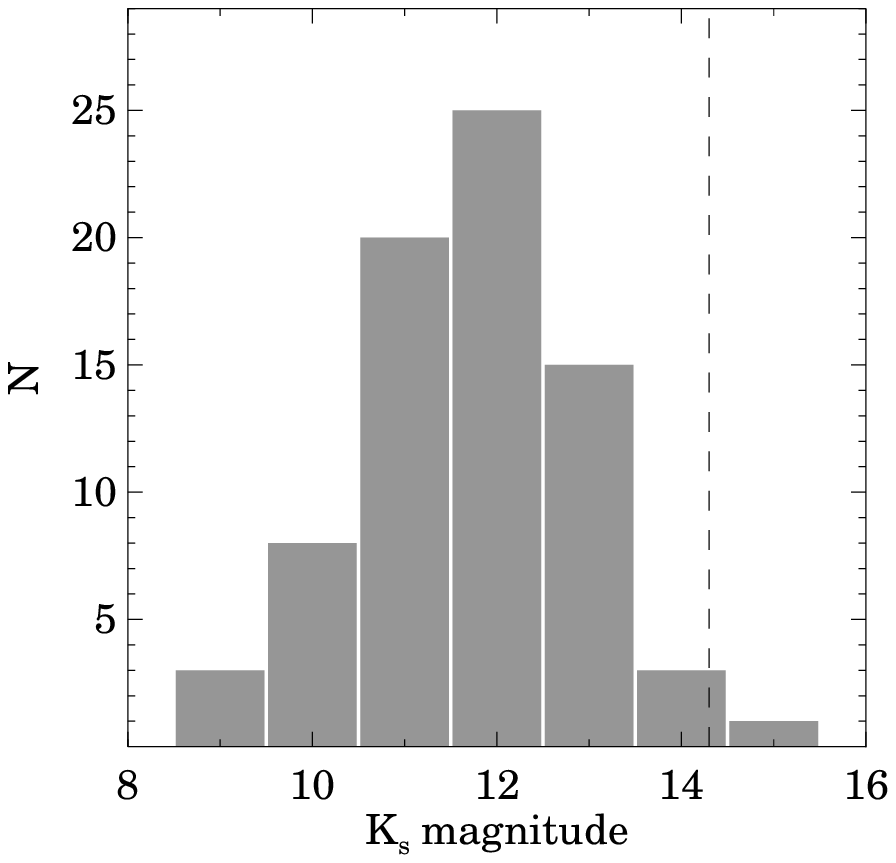}}
\caption{Histogram of the \ha\ equivalent widths (left) and 2MASS {\it K\/}$_s$ magnitudes of the \ha\ emission stars. The vertical dashed line indicates the 2MASS K$_\mathrm{s}$ limiting magnitude.}
\label{Fig_hist1}
\end{figure}

The \ha\ emission stars are candidate classical T~Tauri stars (CTTSs) born in Lynds~1340. Twelve target stars of the longslit observations exhibited emission spectra characteristic of classical T~Tauri stars. In addition to eight [KOS94]\,HA stars we detected \ha\ emission in the spectra of three stars associated with the RNO~8 nebulosity, and in that of a faint star next to the F4 type pre-main sequence star RNO~9, referred to as RNO~9\,B in Table~\ref{Table_splog}. The derived spectral types and measured equivalent widths of the \ha\ line are shown in Table~\ref{Table_spres}. Ten of these stars are located within the boundaries of the WFGS2 observations, and were detected as \ha\ emission stars in the WFGS2 images. To compare the equivalent widths measured with different methods, we list the EW(\ha), measured in the WFGS2 spectra of these stars, in the last column of Table~\ref{Table_spres}. The comparison suggests no systematic difference between the equivalent widths measured in the slitless and longslit spectra.
The left panel of  Fig.~\ref{Fig_hist1} shows the histogram of the EW(\ha) of the 77 emission line stars, detected during the WFGS2 and longslit observations. Most of the measured equivalent widths are between 10 and 100\,\AA, typical for classical T~Tauri stars \citep[e.g][]{Fernandez95,Reipurth96}. These strong emission lines originate from magnetospheric accretion and accretion-related winds \citep[e.g.][]{Muzerolle98}. The highest value (496\,\AA) belongs to the eruptive star V1180~Cas \citep{Kun2011}, and the next highest one (280\,\AA) was detected in HH\,488\,S \citep{KAY03}. EW(\ha)$ < 10$\,\AA\ were measured in the spectra of not more than two stars. The histogram of the \ks\ magnitudes of the \ha\ emission stars is displayed in the right panel of Fig.~\ref{Fig_hist1}. 

\placefigure{Fig_hist1}

\subsubsection{Spectral Energy Distributions}
SEDs of the \ha\ emission stars are shown in Fig.~\ref{Fig_sedha}. The dereddened SED and that of the best fitting photosphere are also plotted. The photometry-based spectral type and extinction, as well as the SED slope type are indicated in each plot. 
According to the classification scheme \citep{Lada,Greene}, our list of \ha\ emission stars contains one Class~I source, 5 Flat SED, 64 Class~II, and 5 Class~III sources. Twenty-three of the Class~II stars have primordial circumstellar disks (II\,P), 32 ones are surrounded by anemic disks (II\,A), and 9 pre-transitional / transitional disks (II\,T) can be found in the sample.

\placefigure{Fig_sedha}

\subsubsection{Hertzsprung--Russell Diagram}
Figure~\ref{Fig_hrd} shows the distribution of the \ha\ emission stars in the HRD. Filled circles indicate the stars with spectral types determined from spectroscopic observations, and open circles show the stars whose spectral types were derived from photometric data. The diagram suggests that most of our selected candidate YSOs are pre-main sequence stars between ages of 1--3 million years, in the mass interval of 0.25--2.0\,\msun, adopting a distance of 825\,pc. Part of the stars are apparently older, and there are a few objects near or even below the ZAMS. The \ha\ star below the ZAMS is 2MASS~02275976+7235561, aka HH\,488\,S. This object is a binary or multiple system \citep{KAY03}, exhibiting a Flat SED (Fig.~\ref{Fig_sedha}). Another underluminous star,  apparently located on the ZAMS, is the central star of RNO~8, mentioned in Sect.~\ref{Sect_imtts}. These stars may have nearly edge-on disks, blocking most of the optical photospheric fluxes. Combination of random uncertainties in photometry, spectral type, extinction,
and veiling may also result in uncertain values of luminosity and temperature \citep[see][]{Manara2013}. We used \citeauthor{siess}'s \citeyearpar{siess} pre-main sequence evolutionary models to estimate the masses of the \ha\ emission stars and their uncertainties. We did not attempt to deduce age distribution from the luminosity distribution. Spectral types, visual extinctions, $T_\mathrm{eff}$, $L$\,/$L_{\sun}$, and the SED slope types of the \ha\ emission stars are listed in Table~\ref{Table_halpha2}. 

\placefigure{Fig_hrd}

\subsubsection{Accretion Rates}
The major source of the \ha\ emission of pre-main sequence stars is gas falling onto the stellar surface along magnetospheric accretion columns. Several empirical relationships have been established between the luminosity of the \ha\ line and accretion luminosity \citep[e.g.][]{Dahm08,HH2008,Fang09,Barentsen}. We computed accretion rates for our \ha\ emission stars using the relationship established by \citet{Barentsen} for the \ha\ emission stars of IC\,1396, spreading nearly on the same mass interval as our stars: 
\begin{equation}
\log (L_\mathrm{acc}/L_{\sun})=(1.13\pm0.07)\log(L_{\ha}/L_{\sun}) +(1.93\pm0.23)\label{eq1} \end{equation}

\noindent where $L_{\ha}$ is the luminosity of the \ha\ emission line. To convert
the luminosity $L_\mathrm{acc}$ to accretion rate $\dot{M}_\mathrm{acc}$, according to the relationship
$$L_\mathrm{acc} \approx \frac{G\dot{M}_{acc} M_{*}}{R_{*}}\left(1-\frac{R_{*}}{R_0}\right),$$
\noindent we took the stellar mass $M_{*}$, and radius $R_{*}$ from the HRD, and adopted $R_{0} \approx 5\,R_{*}$ for the inner radius of the gaseous disk \citep{Gullbring}. Logarithms of the resulting $\dot{M}_\mathrm{acc}$ accretion rates are listed in the eighth column of Table~\ref{Table_halpha2}, and $\dot{M}_\mathrm{acc}$ is plotted against the stellar mass $M_{*}$ in Fig.~\ref{Fig_mdot}. Stars of different SED shapes are distinguished. Filled circle indicates the only Class~I \ha\ star, and open circles mark the Flat SED sources. Squares show the II\,P type SEDs. Upward triangles are for II\,A type SEDs, and downward triangles mark the II\,T type SEDs. Class~III sources are plotted with diamonds. Similarly to several other young stellar groups \citep[e.g][]{NPR2006,HH2008,Barentsen}, a trend can be seen in the widely scattered points. The linear fit to the data, 
$$\log \dot{M} = (1.49 \pm 0.36) \log M - (7.87 \pm 0.09),$$ 
\noindent is shown by the solid line. Its slope is consistent with the values between 1.0 and 3.0, found for other star-forming regions \citep[e.g][]{NPR2006,HH2008,Barentsen}. The median $\dot{M}_\mathrm{acc}$ of the sample is $7.6\times10^{-9}$\,M$_{\sun}$\,yr$^{-1}$, a typical value for T~Tauri stars.
Taking into account the 10\% uncertainty of EW(\ha) and the errors listed in Table~\ref{Table_halpha2} we find that the accretion rates are accurate within $\approx 0.3--0.5$~order of magnitude. Using another empirical relationship results in slightly higher accretion luminosities, but does not affect the shape of the $\dot{M}_\mathrm{acc}$ vs. $M_\mathrm{star}$ plot.

\placefigure{Fig_mdot}

  The {\it U\/}-band magnitudes, available for the H$\alpha$ emission stars, may offer an independent way to derive accretion rates. Since most of our stars are faint in the {\it U\/}-band, and their spectral types, determined from non-simultaneous photometric data, can be regarded as preliminary estimates, we calculated the accretion rates for a carefully  subsample. First we selected stars with {\it U\/}$ < 21.0$~mag and $\delta U < 0.1$. Then we inspected the SEDs of the pre-selected stars and omitted those whose SED indicated photometric variation between the {\it I\/}-band (obtained in 2005) and {\it J\/}-band (1999) fluxes, adding to the uncertainties of the derived spectral type and extinction. For the fifteen remaining stars (Nos. 1, 7, 9, 16, 18, 20, 28, 29, 32, 41, 43, 46, 47, 65, 66) we computed the {\it U\/}-band excess luminosities and transformed them into accretion luminosities following the procedures described by \citet{Gullbring}. Photospheric {\it U\/}-band luminosities were adopted from \citet{KH95}. Comparing the result with the \ha\ luminosities we obtain the relation
$$\log (L_\mathrm{acc}/L_{\sun})=(1.18\pm0.11)\log(L_{\ha}/L_{\sun}) +(1.65\pm0.22),$$
\noindent compatible with Eq.~\ref{eq1}. Figure~\ref{Fig_uha} shows the relation between $\log L_{\ha}$ and $\log L_\mathrm{U_{exc}}$. The estimated uncertainities are within an order of magnitude. 

\placefigure{Fig_uha}

\section{DISCUSSION}
\label{Sect_5}

\subsection{New Features of the Cloud Structure} 
The new extinction map of L1340 reveals a shallow, strongly fragmented cloud. The average hydrogen column density within the  $A_\mathrm{V} = 1$~mag contour is $N(H_2) \approx 1.87\times10^{21}$\,cm$^{-2}$. If they exist, high column density regions, a prerequisite of star formation, are smaller in diameter than some 0.35~pc, the resolution of our map. The 850~\mum\ map of L1340\,B \citep{DiFran} and the NH$_3$ cores (Paper~II) point to such regions. The average hydrogen column densities, obtained from $A_\mathrm{V}$ for the clumps L1340\,A, B, and C, are around $N(H_2) \la 2.5\times10^{21}$\,cm$^{-2}$ (Table~\ref{Table_clump}), lower than the detection threshold for C$^{18}$O \citep[e.g.][]{Pineda08}, suggesting that C$^{18}$O was detected from dense, small regions of a knotty formation. The past and present winds and outflows of embedded intermediate-mass stars have probably played a role in sustaining the knotty structure, as it has been discussed by \citet{Offner2015}. This effect is demonstrated by Fig.~\ref{Fig_zoom}, in which the extinction map of the central part of the clump L1340\,B, containing the stars illuminating the DG\,9 nebula is displayed, together with the 850-\micron\ contours. The figure reveals several small, bubble-like features in the extinction, the most conspicuous one around the A0-type star 02290319+7259366 (star I5 in Fig.~\ref{Fig_sptarget}), bordered by dark lanes or submillimeter emission and embedded protostellar sources. 

\placefigure{Fig_zoom}

\subsection{The Optically Selected YSO Population of L1340} 
The pre-main sequence evolutionary time of the most massive star associated with L1340 is a few times $10^5$~years \citep{PS93,siess}. In spite of the absence of conspiuous pre-main sequence features this star may be very young, similar in several respects to BD$+65^{o}1638$ in the young cluster NGC\,7129 \citep{DH2015}. The  temporary, weak emission component, detected in its \ha\ line, supports this argument. The weak emission components, suspected in the \ha\ lines of other intermediate-mass stars, have to be confirmed by high-resolution spectroscopic observations. 

Our sample of candidate T~Tauri stars contains 5 G-type, 50 K-type, and 21 M-type stars. The median mass of the \ha\ emission population is 0.7\,M$_{\sun}$. A sizeable part of the young population of the cloud is expected below the detection threshold of the slitless spectroscopic observations. The lower mass limit is 0.25~M$_{\sun}$ corresponding to a spectral type of $\sim$~M3. The cumulative distribution of the derived stellar masses is shown in Fig.~\ref{Fig_mf}. The slope of the line, fitted to the data over the $0.3 \leq M_{*}/M_{\sun} \leq 3.0$ region,
 $$\log N(M_{*} > m) = (-1.38 \pm 0.09) \log m + (1.44 \pm 0.025) $$
\noindent is compatible with that of Salpeter's \citeyearpar{Salpeter} initial mass function.

\placefigure{Fig_mf}

The surface distribution of the candidate young stellar population, overplotted on the extinction map of the cloud, is displayed in Fig.~\ref{Fig_map}. G, K, and M-type T~Tauri star candidates are plotted with different colors, and the SED classes are distinguished by the plotting symbol. Filled symbols indicate stars with accretion rates higher than the median $7.6\times10^{-9}$\,M$_{\sun}$\,yr$^{-1}$, while open symbols show the stars accreting more slowly than the median. Figure~\ref{Fig_map} shows that the \ha\ emission stars are scattered over the observed area. [KOS94]\,HA~4 and [KOS94]\,HA~13 are projected outside of the area covered by the WFGS2 observatons, suggesting a possible more widely extended pre-main sequence star population. 

\placefigure{Fig_map}

A striking group is the cluster RNO\,7, projected within the clump L1340\,A, and containing 14 \ha\ emission stars. Further small groups, consisting of less than ten members can be seen over the observed area. The most massive stars, including four of the five G-type T~Tauri stars, are associated with the most massive cloud clump L1340\,B. \co\ and NH$_3$ observations (Paper~I, Paper~II) have shown that the kinetic temperature of the molecular gas is higher in clump~B than in clumps A and C. The stars projected on L1340\,B, however, form a widely scattered aggregate, without conspicuous clustering, unlike other star-forming regions, exhibiting stronger clustering around more massive young stars  \citep[e.g.][]{Testi99,Lund14}. 

Figure~\ref{Fig_map} shows clearly that K-type stars dominate our \ha\ emission sample. It suggests also that II\,T type disks tend to be projected on low-extinction regions. Within the wide scatter of the points in Fig.~\ref{Fig_mdot} it can be noticed that none of the downward triangles (pre/transitional disks) are found above the fitted line, indicating that their accretion rates are lower than the average. To examine in more detail how the shape of the SED of a \ha\ emission star is related to stellar properties, accretion activity, and location within the cloud, we show in Table~\ref{Table_sedcomp} the average EW(\ha), $A_\mathrm{V}$, $\log T_\mathrm{eff}$, $\log L$, and $\dot{M}_\mathrm{acc}$ for each SED class and subclass. The comparison shows that Flat and Class II\,P sources are located in regions of higher extinction, they are more massive and their accretion rates are higher than those of the stars possessing II\,A and II\,T type disks. Similar correlations of the SED shapes with accretion rates and stellar distributions were reported for the Taurus Class~II stars by \citet{Najita2007} and \citet{Luhman2010}, respectively. 
It is noteworthy that the transitional disks of our sample (stars 7, 50), exhibiting photospheric SED below 24\,\micron, and weak/pre-transitional disks without excess emission below 8\,\micron\  (stars 3, 4, 6) were identified as \ha\ emission stars, suggesting CTTS-like accretion rates.

\placetable{Table_sedcomp}

The age range of our target stars can be poorly constrained without further spectroscopic data. The HRD positions suggest that it may be comparable with that of the Taurus pre-main sequence population \citep[cf. fig.~7 of][]{Luhman2009}. The number of Class~II YSOs are also similar in both star-forming regions \citep{Luhman2010,Kun2016}, whereas they are different in volume, molecular mass and kinetic temperature, structure, and Galactic environment \citep{Kenyon2008}. To check whether disk properties observed in the two apparently coeval star-forming regions can be distinguished or not, we collected statistical data in Table~\ref{Table_taucomp}.  We keep in mind that L1340 is some six times as distant as Taurus, therefore the low-mass side of its mass spectrum is more incompletely sampled. Furthermore, the \ha\ emission stars are more massive and stronger accretors than the Class~II average of the region. Table~\ref{Table_taucomp} suggests that the Class~II disks of L1340 and Taurus cannot be distinguished based on \textit{Spitzer\/} color indices and mean accretion rates.

\placetable{Table_taucomp}

\subsection{Comparison with Other IM\,SFRs}
\citet{Lund14,Lund15} identified and studied a large sample of Galactic IM\,SFRs. Most of them are located significantly farther from us than L1340 and typically contain loose clusters with less than 100 members. The mean gas column densities and $^{13}$CO linewidths observed for L1340 are within the range found for this sample. The sample of 36 young clusters within our 1-kpc environment, studied by \citet{Gutermuth09}, contains several young stellar groups whose most massive star is around 5--6\,M$_{\sun}$ (e.g. IC\,348, IRAS\,20050+2720, BD$+40^\mathrm{o}4124$, NGC~7129). Unlike L1340, most of these star-forming regions are parts of giant molecular cloud complexes. 
L1340 is located at some 160~pc above the Galactic plane, near the outermost boundary of the molecular gas disk, far from any known giant molecular cloud. In this respect it is similar to NGC\,7129, lying at a similar distance from the Galactic plane. Molecular clouds and star formation at intermediate Galactic latitudes may be produced by expanding superbubbles. No such object has been identified in the environment of L1340. Infalling high velocity clouds, or Kelvin--Helmholtz instabilities arising at the shearing surface between gas layers of different velocities may also compress the gas. Detailed examination of the velocity fields of the apparent swirling structures, seen in the \textit{WSSA\/} \citep{Meisner14} image of the large-scale environment of L1340 may shed light on the interstellar processes leading to star formation in L1340.

\section{CONCLUSIONS}
\label{Sect_6}

We studied the structure and optically selected young stellar population of the molecular cloud L1340, located within our 1-kpc environment, but poorly studied so far. 
Our optical spectroscopic and photometric search for young stellar objects associated with the molecular cloud L1340 revealed that the most massive star associated with this cloud is a B4 type, 5\,\msun\ star. The new spectroscopic and photometric data of the intermediate mass members led to a distance of $825^{+110}_{-80}$\,pc, and revealed 14 candidate members of the young stellar population with $M \gtrsim 2$\,\msun. Our search for \ha\ emission line stars, conducted with the \textit{WFGS2\/} instrument on the $2.2$-meter telescope of the University of Hawaii and covering a $30\arcmin \times 40\arcmin$ area, resulted in the detection of 75 candidate low-mass pre-main sequence stars, 58 of which are new. We constructed SEDs of our target stars, based on \sdss, \tm, \spitzer, and \wise\ photometric data, derived their spectral types, extinctions, and luminosities from {\it BVRIJ\/} fluxes, estimated masses by means of pre-main sequence evolutionary models, and examined the disk shapes utilizing the 2--24\,\micron\ interval of the SED. We measured the equivalent width of the \ha\ line and derived accretion rates. The new extinction map of L1340, based on \sdss\ data, suggests a shallow cloud of clumpy structure, having a mass of some 3700~M$_{\sun}$. The optically selected sample of pre-main sequence stars has a median effective temperature of 3970\,K, stellar mass 0.7\,M$_{\sun}$, and accretion rate of 7.6$\times10^{-9}$~\msun\,yr$^{-1}$. 
The highest mass stars and the highest extinction are associated with the largest clump of the cloud. However, the surface distribution of the young stars is more scattered in the largest clump than in the smaller ones.

\begin{acknowledgements}
Our results are based on observations with the $2.2$-m telescope of the University of Hawaii and we thank Colin Aspin and Mark Willman for their interest and support. This work makes use of observations made with the \textit{Spitzer Space Telescope}, which is operated by the Jet Propulsion Laboratory, California Institute of Technology under a contract with NASA. This research utilized observations collected at the Centro Astron\'omico Hispano Alem\'an (CAHA) at Calar Alto, operated jointly by the Max-Planck Institut f\"ur Astronomie and the Instituto de Astrof\'{\i}sica de Andaluc\'{\i}a (CSIC). We are grateful to G\'abor F\H{u}r\'esz for observing the FAST spectra.
This work makes use of data products from the Wide-field Infrared Survey Explorer, which is a joint project of the University of California, Los Angeles, and the Jet Propulsion Laboratory/California Institute of Technology, funded by the National Aeronautics and Space Administration. This research has made use of the NASA/ IPAC Infrared Science Archive, which is operated by the Jet Propulsion Laboratory, California Institute of Technology, under contract with the National Aeronautics and Space Administration. This research has made use of the VizieR catalogue access tool, CDS, Strasbourg, France. The original description of the VizieR service was published in A\&AS 143, 23.
This work also makes use of data products from the SDSS-II data. Funding for SDSS-II has been provided by the Alfred P. Sloan Foundation, the Participating Institutions, the National Science Foundation, and the U.S. Department of Energy Office of Science. The SDSS  web site is http://www.sdss.org/. Financial support from the Hungarian OTKA grants K81966 and K101393 is acknowledged. This work was partly supported by the Momentum grant of the MTA CSFK Lend\"ulet Disk Research Group.
\end{acknowledgements}

\appendix

\section{PHOTOMETRIC DATA OF THE OPTICALLY SELECTED CANDIDATE PMS STARS}
 We list {\it UBVR}$_\mathrm{C}${\it I}$_\mathrm{C}$ and \textit{2MASS\/} \textit{JHK$_s$} magnitudes of the optically selected candidate young stars associated with L1340 in Table~A1. Table~A2 contains the \textit{Spitzer\/} [3.6], [4.5], [5.8], [8.0], [24], and [70.0] magnitudes, and \textit{AllWISE\/} [3.4], [4.6], [12], and [22] magnitudes for the same stars.

\clearpage

\newpage

\newpage

\clearpage

\begin{figure*}
\centering{\includegraphics[width=16cm]{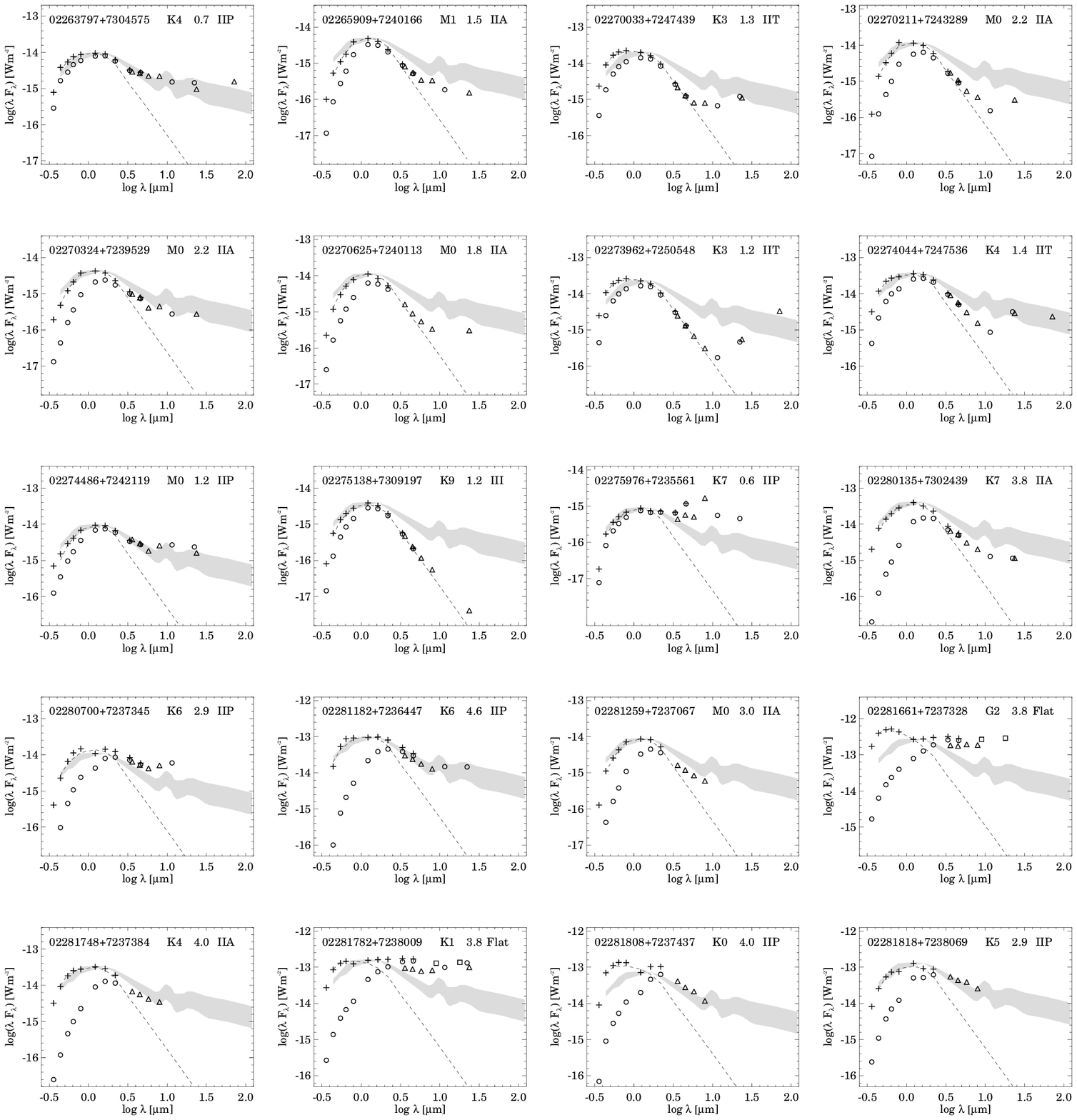}}
\caption{SEDs of the \ha\ emission stars. Open circles show the \textit{SDSS}, \textit{2MASS}, and \textit{WISE} data, and triangles result from our \spitzer\ photometry. Open squares indicate \textit{Akari\/} IRC fluxes. Plusses indicate the dereddened SED, and the dashed line shows the photospheric SED of the spectral type, obtained by fitting a model to the data (see the text). The gray shaded area indicates the median SED of the T~Tauri stars of the Taurus star-forming region \citep{DAlessio}. \textit{2MASS} identifiers, spectral types, and $A_\mathrm{V}$ extinctions, derived from photometric data, as well as the SED class are indicated at the top of each panel.}
\label{Fig_sedha}
\end{figure*}

\begin{figure*}
\addtocounter{figure}{-1}
\centering{\includegraphics[width=16cm]{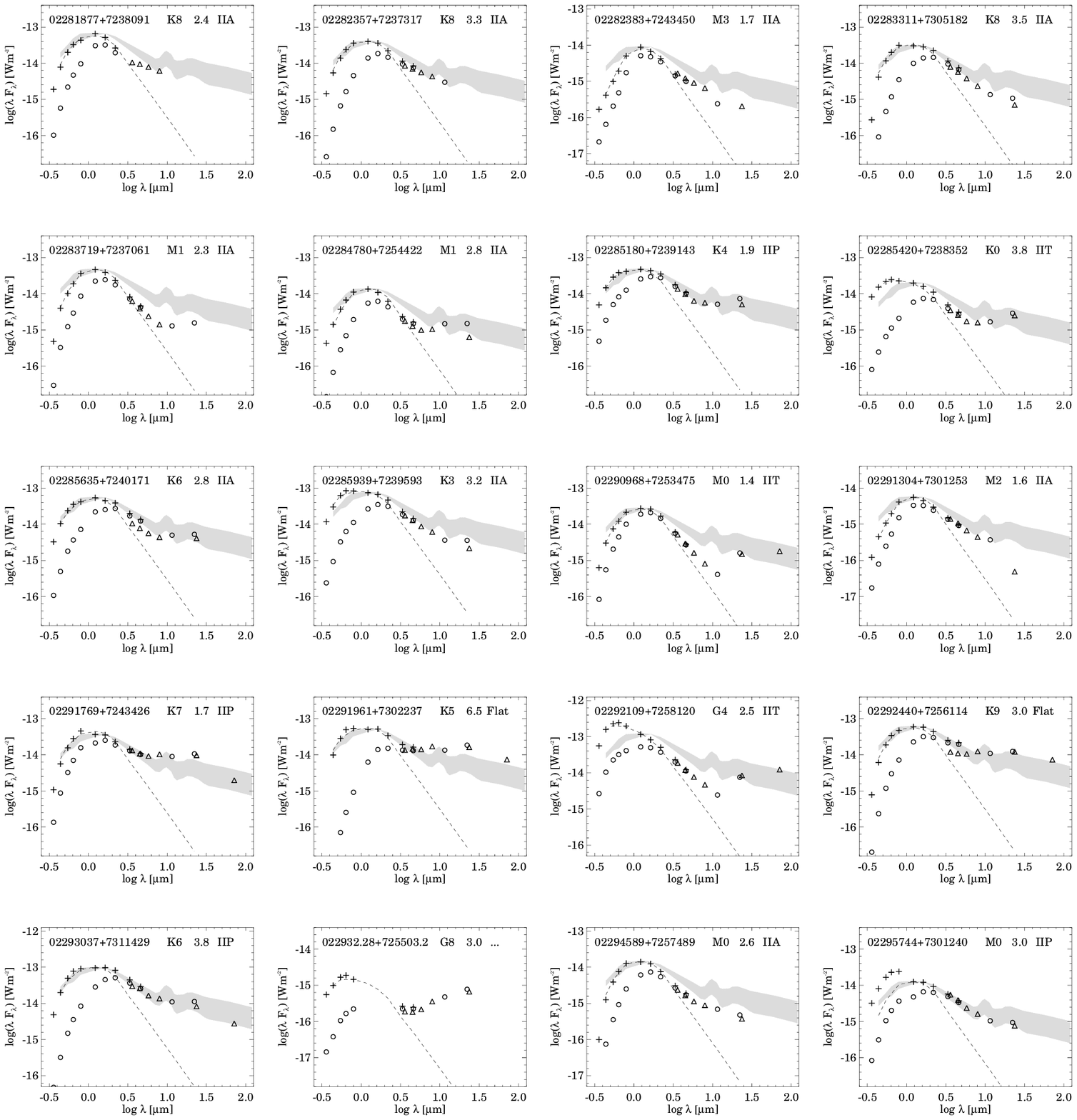}}
\caption{SEDs of the \ha\ emission stars (continued.)}
\label{Fig_sedha}
\end{figure*}

\begin{figure*}
\addtocounter{figure}{-1}
\centering{\includegraphics[width=16cm]{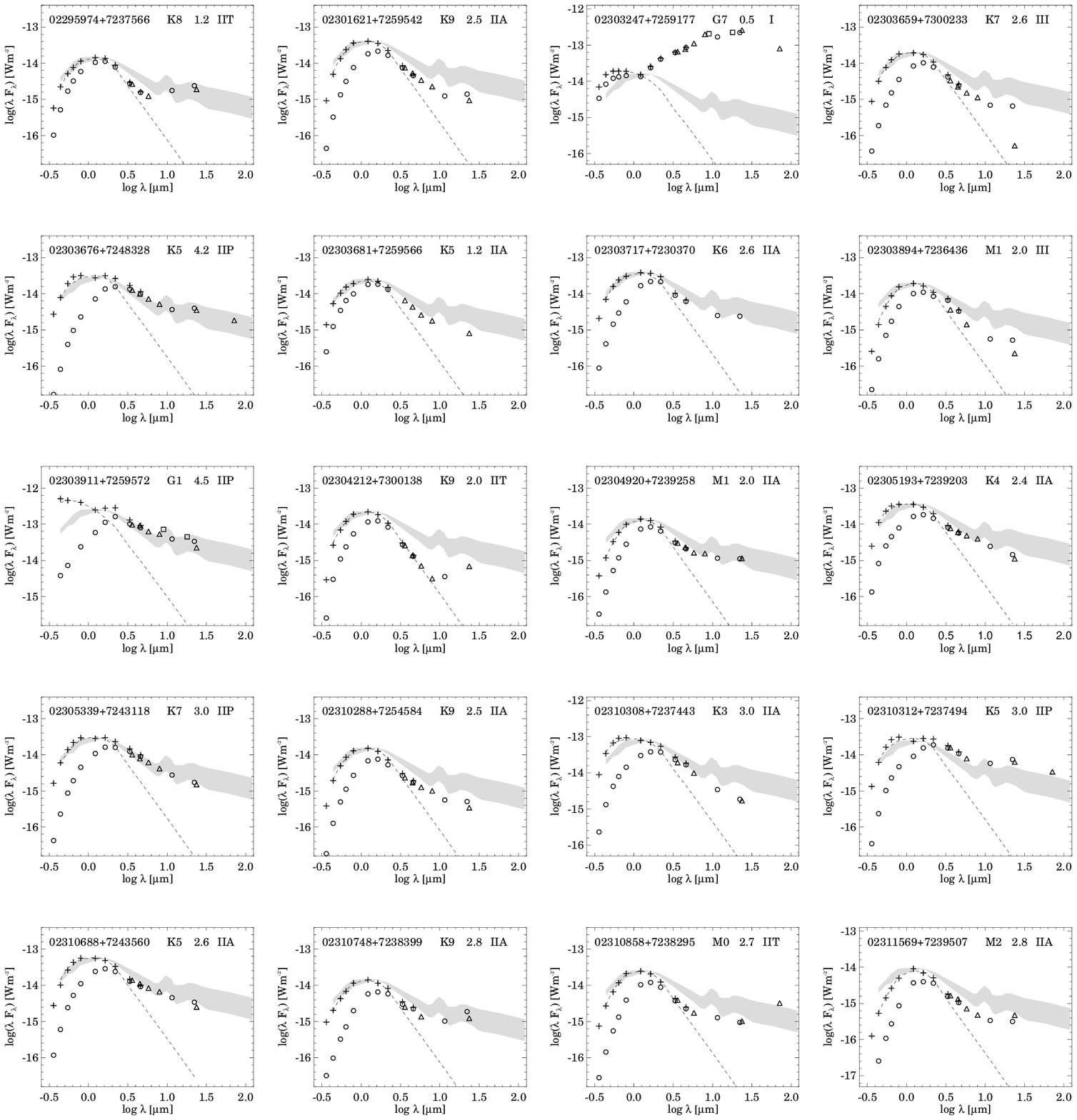}}
\caption{SEDs of the \ha\ emission stars (continued.)}
\label{Fig_sedha}
\end{figure*}

\begin{figure*}
\addtocounter{figure}{-1}
\centering{\includegraphics[width=16cm]{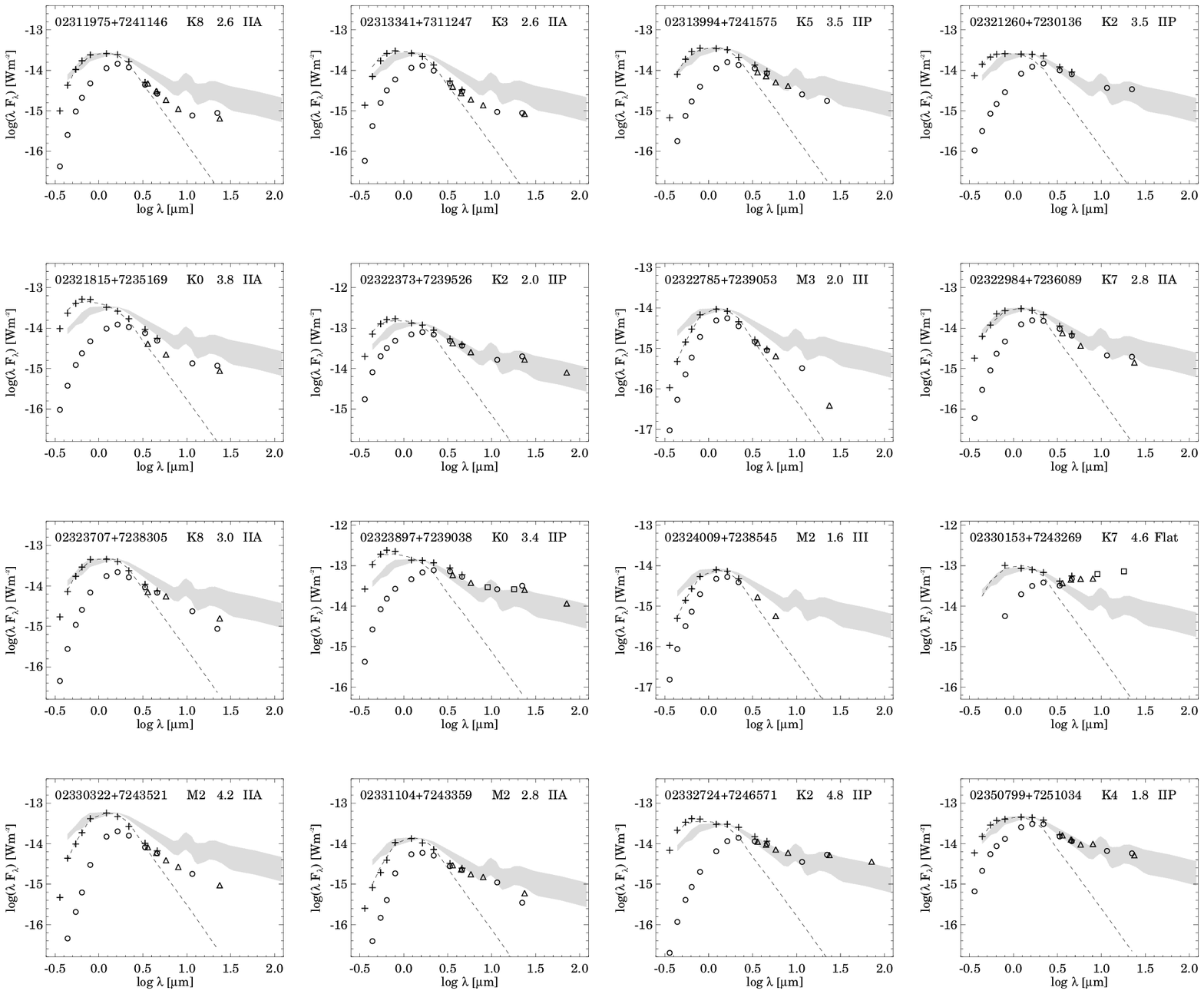}}
\caption{SEDs of the \ha\ emission stars (continued.)}
\label{Fig_sedha}
\end{figure*}

\clearpage

\begin{figure*}
\centering{\includegraphics[width=16cm]{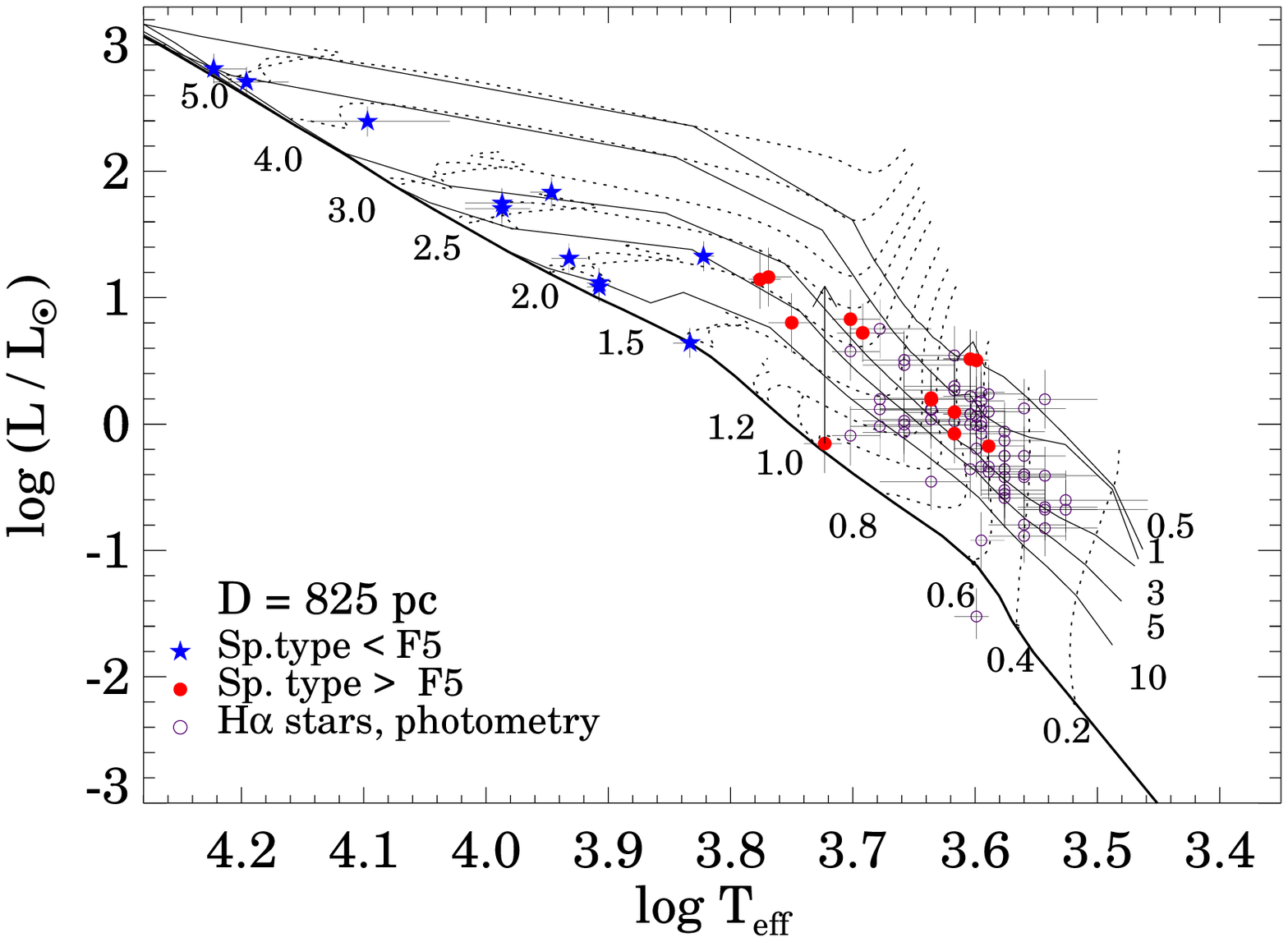}}
\caption{Hertzsprung--Russell diagram of the candidate young stellar population of L1340. Blue star symbols indicate the stars earlier than F5, red filled circles show those later than F5, listed in Table~\ref{Table_splog}. Open circles indicate sources whose spectral types were derived from photometric data. Evolutionary tracks (dotted lines) and 0.5, 1, 3, 5, 10 million year isochrones (thin solid lines), and the main sequence (thick solid line) are from \citet{siess}. The stellar bolometric luminosities were calculated for a distance of 825\,pc.}
\label{Fig_hrd}
\end{figure*}

\begin{figure}
\centering{\includegraphics[width=8cm]{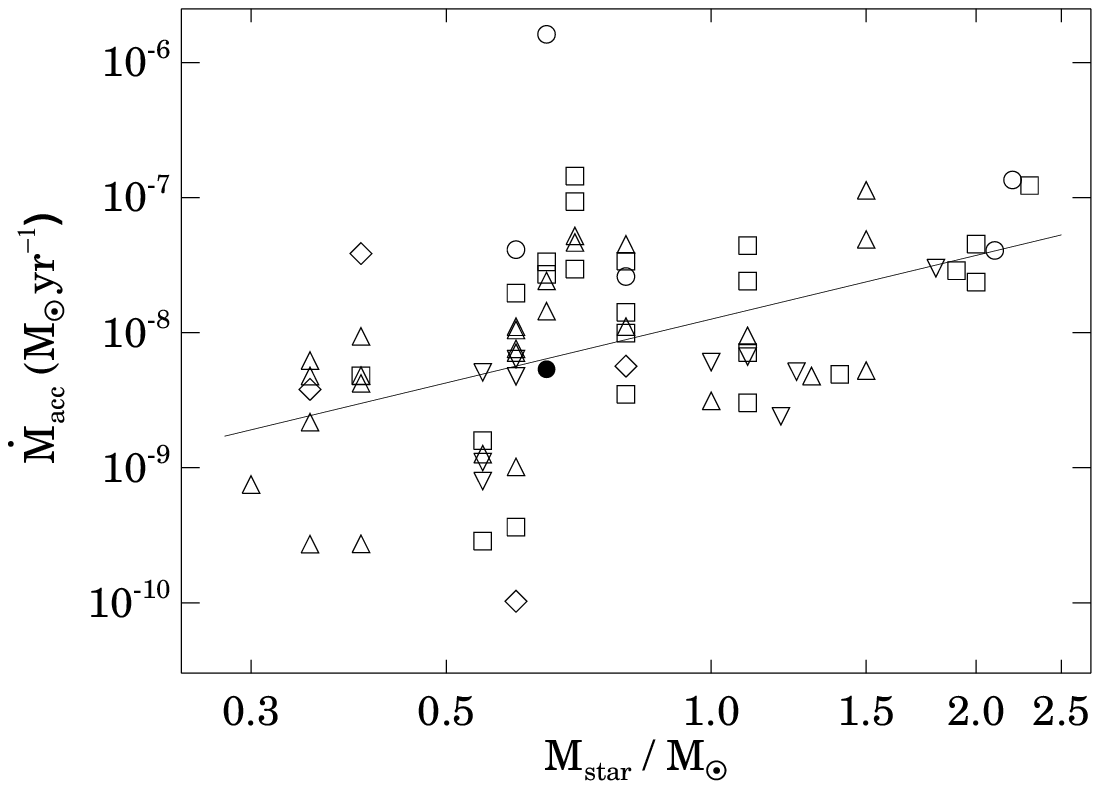}}
\caption{Accretion rates, derived from the \ha\ luminosites as a function of the stellar mass. The filled circle is the only  Class~I source, open circles indicate the Flat SED sources. II\,P type SEDs are marked by squares. Upward triangles show II\,A sources, downward triangles are for II\,T type SEDs, and diamonds show the Class~III young stars. The linear fit to the data is also drawn.}
\label{Fig_mdot}
\end{figure}

\begin{figure}
\centering{\includegraphics[width=8cm]{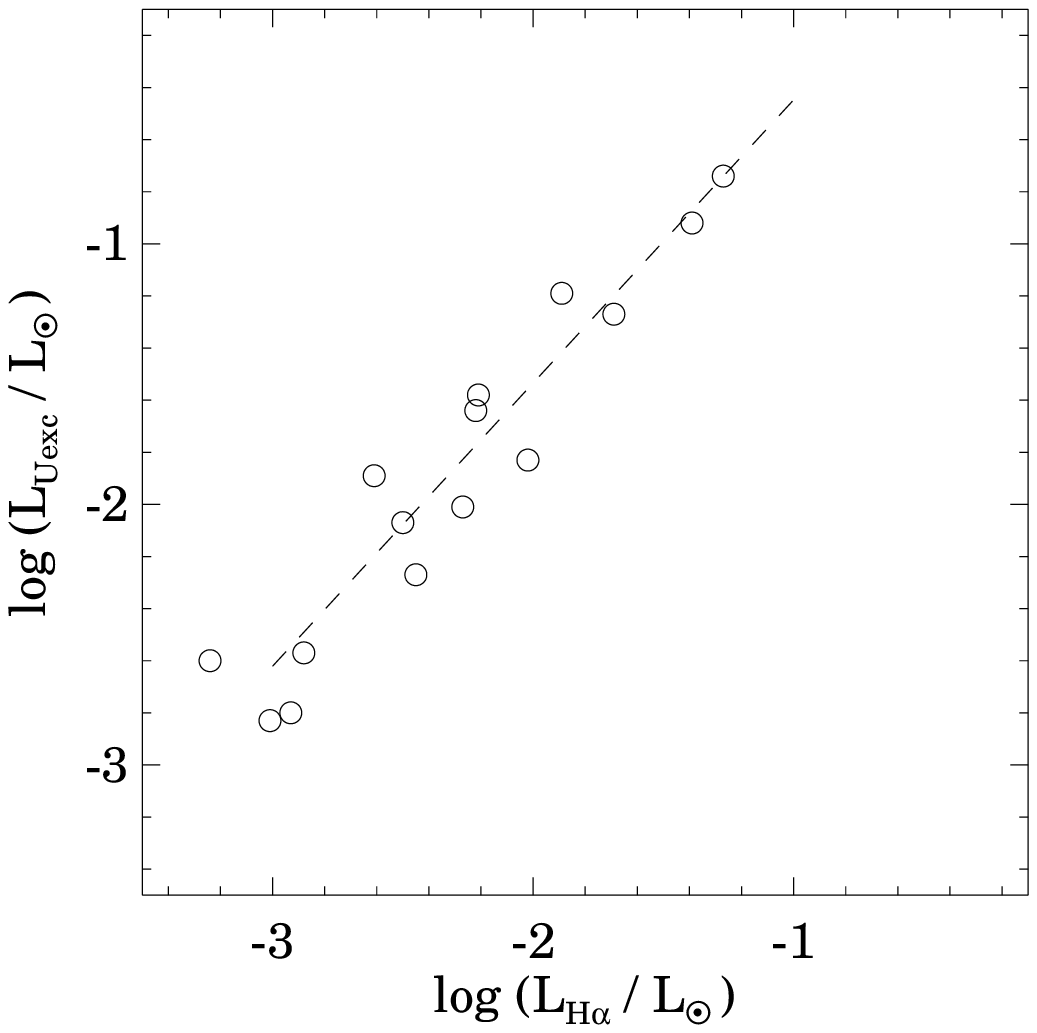}}
\caption{Logarithms of {\it U\/}-band excess luminosities of fifteen selected \ha\ emission stars, plotted against the logarithms of \ha\ luminosities.}
\label{Fig_uha}
\end{figure}

\begin{figure}
\centering{\includegraphics[width=8cm]{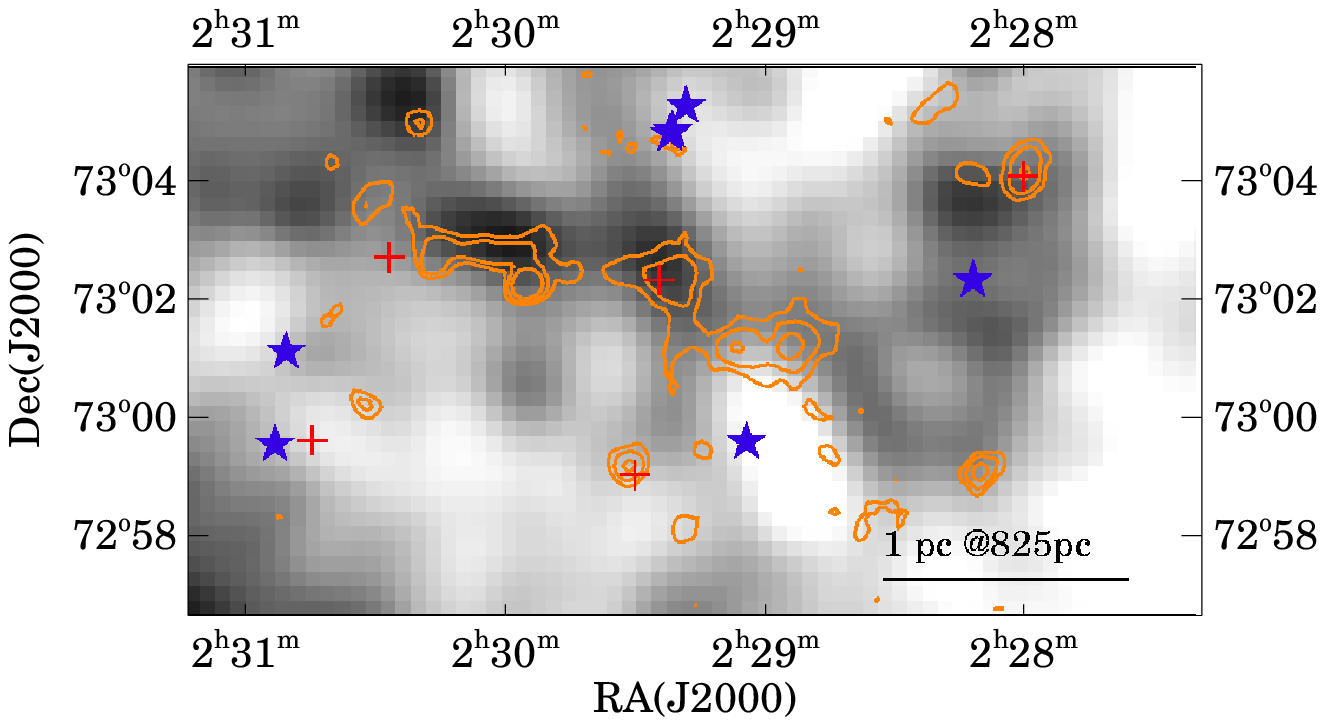}}
\caption{Extinction map of the central $\sim 18\times11$~arcmin area of the clump L1340\,B,  with the 850\,\mum\ contours (orange) overplotted. Blue asterisks indicate the stars associated with the reflection  nebula DG\,9, and red plusses show the \textit{IRAS} point sources.}
\label{Fig_zoom}
\end{figure}

\begin{figure}
\centering{\includegraphics[width=8cm]{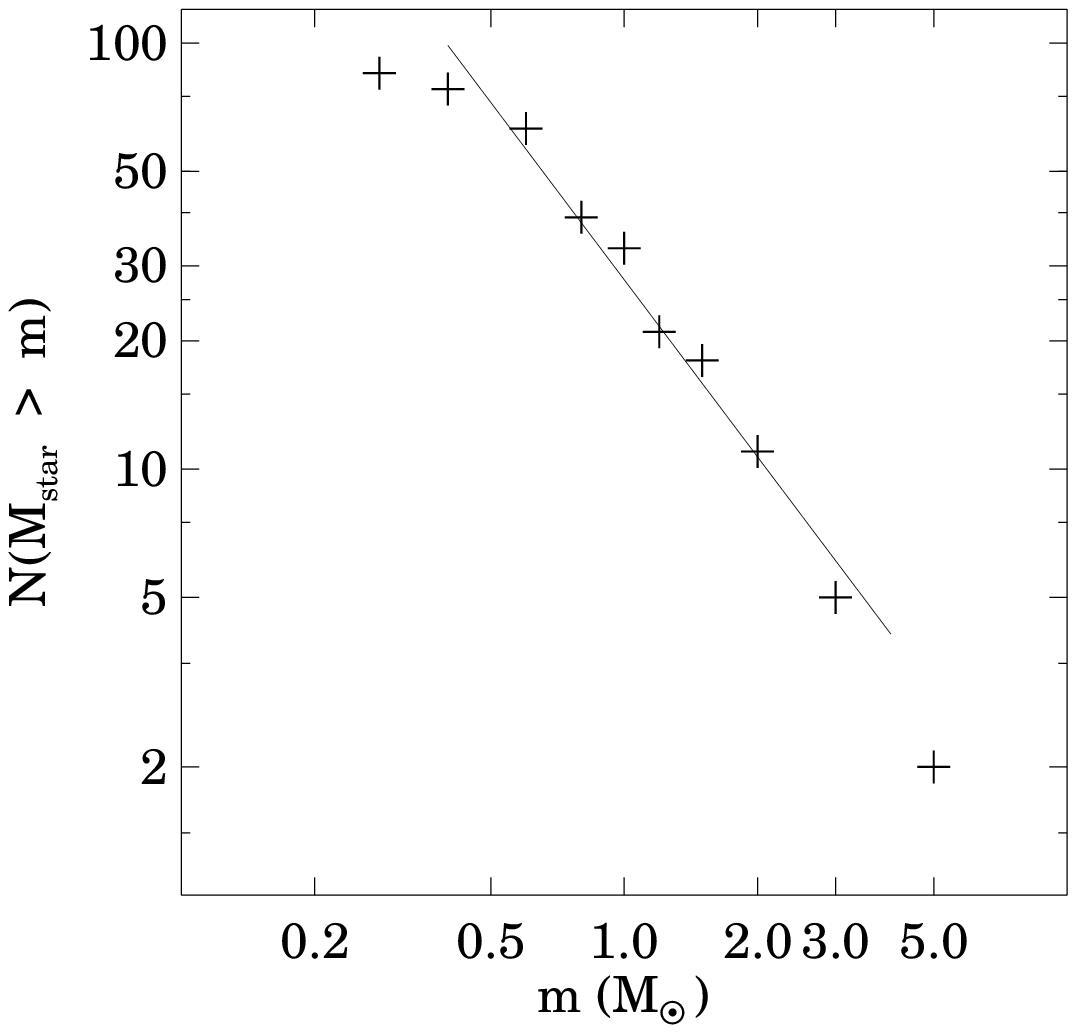}}
\caption{Cumulative mass distribution of the candidate young stars identified during the present work, derived from their positions in the HRD, using the pre-main sequence evolutionary models of \citet{siess}. Solid line shows the linear fit to the data between 0.5 and 3\,M$_{\sun}$. }
\label{Fig_mf}
\end{figure}

\begin{figure*}
\centering{\includegraphics[width=15cm]{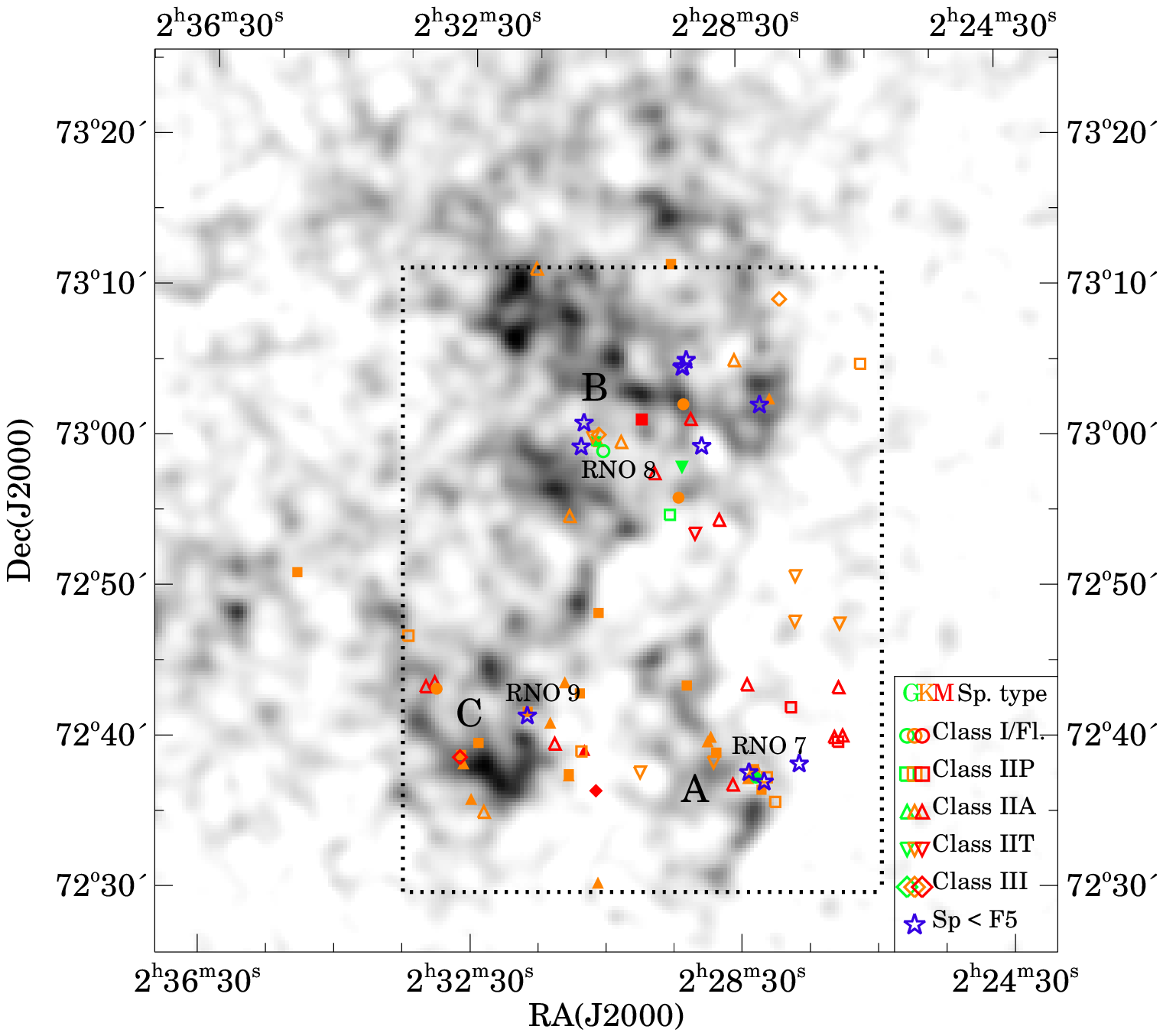}}
\caption{Distribution of the intermediate-mass stars earlier than F5 type (blue star symbols) and \ha\ emission stars, overplotted on the extinction map of the region. Dotted lines outline the area of the WFGS2 survey. The lowest contour is drawn at $A_\mathrm{V}=1.5$~mag, and the increment is 1~mag. Filled symbols indicate stars with $\dot{M}_\mathrm{acc} \ge 7.6\times10^{-9}$\,M$_{\sun}$\,yr$^{-1}$, and stars plotted with open symbols accrete slower than the median of the sample.}
\label{Fig_map}
\end{figure*}

\newpage

\begin{deluxetable}{rccccl}
\tabletypesize{\scriptsize}
\tablecolumns{6}
\tablewidth{0cm}
\tablecaption{Log of spectroscopic observations \label{Table_splog}}
\tablehead{
\colhead{No.} & \colhead{2MASS Id} & \colhead{Other id.} &  \colhead{Date of obs.} & \colhead{$\lambda\lambda$ (\AA)} & \colhead{Tel./Instr.} }
\startdata  
\cutinhead{Candidate Intermediate-mass Young Stars}
 I1 & 02273807+7238267 &  \nodata      & 2011.09.27  & 3800--7500 &  Konkoly RCC    \\  
 I2 & 02280942+7237162\tablenotemark{a} &  \nodata   & 2004.12.08  & 3800--7500 &  FLWO 1.5-m/FAST \\
 I3 & 02281033+7302197 &  $[KOS94]$ R1    & 2004.10.31  & 3800--7500 &  FLWO 1.5-m/FAST     \\  
 I4 & 02282291+7237539\tablenotemark{a} &  \nodata   & 2004.11.06  & 3800--7500 &  FLWO 1.5-m/FAST	   \\
 I5 & 02290319+7259366 &  $[KOS94]$ R2    & 1999.11.05  & 3800--7500 &  FLWO 1.5-m/FAST      \\  
 I6 & 02291684+7305197 &  $[KOS94]$ R3    & 1999.08.07  & 4900--7800 & CA 2.2-m/CAFOS/G-100  \\  
 I6 & 02291684+7305197 &  $[KOS94]$ R3    & 2004.11.06  & 3800--7500 &  FLWO 1.5-m/FAST   \\  
 I7 & 02292000+7304545 &  $[KOS94]$ R3-2A & 2004.11.05  & 3800--7500 &  FLWO 1.5-m/FAST   \\  
 I8 & 02292062+7304514 &  $[KOS94]$ R3-2B & 2004.11.05  & 3800--7500 &  FLWO 1.5-m/FAST \\
 I9 & 02304997+7301100 &  $[KOS94]$ R4    & 2005.09.13  & 4900--7800 & CA 2.2-m/CAFOS/G-100   \\  
 I10 & 02305176+7259481 &  \nodata   & 2004.12.08  & 3800--7500 &  FLWO 1.5-m/FAST	   \\ 
 I11 & 02314031+7241419 &  RNO 9       & 1999.08.07  & 4900--7800 & CA 2.2-m/CAFOS/G-100   \\ 
 I11 & 02314031+7241419 &  RNO 9       & 2004.12.11  & 3800--7500 &  FLWO 1.5-m/FAST  \\ 
\cutinhead{Candidate Low-mass Pre-Main Sequence Stars}
 T1 & 02281661+7237328 &  $[KOS94]$ HA 1  & 1999.08.07  & 4900--7800 &  CA 2.2-m/CAFOS/G-100   \\  
 T2 & 02281782+7238009 &  $[KOS94]$ HA 2  & 1999.08.07  & 4900--7800 & CA 2.2-m/CAFOS/G-100    \\  
 T3 & 02285180+7239143 &  $[KOS94]$ HA 5  & 1999.08.07  & 4900--7800 &  CA 2.2-m/CAFOS/G-100  \\  
 T4 & 02292109+7258120 &  $[KOS94]$ HA 9  & 1999.08.07  & 4900--7800 &  CA 2.2-m/CAFOS/G-100  \\  
 T5 & 02293037+7311429 &  $[KOS94]$ HA 4  & 1999.08.07  & 4900--7800 & CA 2.2-m/CAFOS/G-100  \\  
 T6 & 02303247+7259177 &  IRAS 02259+7246 & 2000.01.05  & 5825--8350 &  NOT/ALFOSC/Grism 8  \\  
 T7 & 02303681+7259566 &  RNO 8\,West	  & 2000.01.05  & 5825--8350 &  NOT/ALFOSC/Grism 8   \\  
 T8 & 02303911+7259572 &  RNO 8\,East	    & 2000.01.05  & 5825--8350 &  NOT/ALFOSC/Grism 8 \\  
 T9 & 02313994+7241575 &  RNO 9\,B	  & 2005.09.13  & 4900--7800 &  CA 2.2-m/CAFOS/G-100   \\  
 T10 & 02323897+7239038 &  $[KOS94]$ HA 10 & 2004.12.10  & 3800--7500 & FLWO 1.5-m/FAST    \\ 
 T10 & 02323897+7239038 &  $[KOS94]$ HA 10 & 2004.12.10  & 3800--7500 & FLWO 1.5-m/FAST    \\ 
 T10 & 02323897+7239038 &  $[KOS94]$ HA 10 & 1999.08.07  & 4900--7800 &  CA 2.2-m/CAFOS/G-100  \\  
 T10 & 02323897+7239038 &  $[KOS94]$ HA 10 & 2004.12.10  & 3800--7500 & FLWO 1.5-m/FAST    \\ 
 T11 & 02330153+7243269 &  $[KOS94]$ HA 11\tablenotemark{b} & 2003.02.05  & 4900--7800 & CA 2.2-m/CAFOS/G-100  \\ 
 T12 & 02350799+7251034 &  $[KOS94]$ HA 13 & 1999.08.07  & 4900--7800 & CA 2.2-m/CAFOS/G-100   \\  
 T12 & 02350799+7251034 &  $[KOS94]$ HA 13 & 2004.12.10  & 3800--7500 & FLWO 1.5-m/FAST    \\ 
\enddata        				  
\tablenotetext{a}{This star was included into the target list due to the blue color indices and projected location within the RNO~7 cluster.}
\tablenotetext{b}{Spectroscopic and photometric variability of [KOS94]~HA\,11 ($\equiv$ V1180~Cas) is described in \citet{Kun2011}}
\end{deluxetable}

\def\ha{\relax \ifmmode {\rm H}\alpha\else H$\alpha$}
\def\ha{\hbox{H$\alpha$}}


\begin{deluxetable}{rlcl}
\tabletypesize{\scriptsize}
\tablecolumns{4}
\tablewidth{0cm}
\tablecaption{Coordinates, H$\alpha$ equivalent widths, and cross-identifiers of the H$\alpha$ emission stars 
 \label{Table_halpha1}}
\tablehead{\colhead{N} & \colhead{2MASS /SDSS Id.} &  \colhead{EW(H$\alpha)$(\AA)}  &  \colhead{Other Id./Associated object}}
\startdata 
  1  &    02263797+7304575 & 84.7$\pm$4.5    &        \\ 
  2  &    02265909+7240166 & 16.7$\pm$2.0    &        \\ 
  3  &    02270033+7247439 & 41.3$\pm$1.3    &        \\ 
  4  &    02270211+7243289 & 15.2$\pm$7.4    &        \\ 
  5  &    02270324+7239529 & 240.0$\pm$80.0  &        \\ 
  6  &    02270625+7240113 & 25.05$\pm$16.8  &  	 \\ 
  7  &    02273962+7250548 & 16.7$\pm$1.1    &         \\ 
  8  &    02274044+7247536 & 24.0$\pm$0.8    &  	\\
  9  &    02274486+7242119 & 37.6$\pm$4.9    &         \\ 
 10  &    02275138+7309197 & 14.0$\pm$2.5    &         \\
 11  &    02275976+7235561 & 280.0$\pm$60.0  &       HH 488S  \\		
 12  &    02280135+7302439 & 171.9$\pm$28.7  &         \\ 
 13  &    02280700+7237345 & 49.0$\pm$5.0    &        RNO7-2  \\
 14  &    02281182+7236447 & 49.4$\pm$4.0    &       RNO7-3  \\ 		
 15  &    02281259+7237067 & \nodata	       &         RNO7-4  \\
 16  &    02281661+7237328 & 18.5$\pm$0.9    &   IRAS 02236+7224, [KOS94]HA 1, RNO7-5, T1  \\     
 17  &    02281748+7237384 & 60.5$\pm$5.0    &   RNO7-6  \\
 18  &    02281782+7238009 & 42.0$\pm$4.2    &  [KOS94]HA 2, RNO7-7, T2  \\    
 19  &    02281808+7237437 & 37.5$\pm$6.0    &  RNO7-8  \\		   
 20  &    02281818+7238069 & 23.6$\pm$2.7    &   RNO7-9  \\ 
 21  &    02281877+7238091 & \nodata	       &  RNO7-10 \\
 22  &    02282357+7237317 & 29.7$\pm$2.4    &   RNO7-11 \\
 23  &    02282383+7243450 & 26.2$\pm$2.0    &    \\
 24  &    02283311+7305182 &  26.9$\pm$4.0   &    \\ 
 25  &    02283719+7237061 & 9.6$\pm$6.8     &   RNO7-12 \\
 26  &    02284780+7254422 & 42.8$\pm$35.2   &   \\ 
 27  &    02285180+7239143 & 84.0$\pm$4.6    &  [KOS94] HA 5, T3  \\	   
 28  &    02285420+7238352 & 68.1$\pm$17.3   &   RNO7-14 \\
 29  &    02285635+7240171 &  83.1$\pm$11.8  &      \\  			
 30  &  022856.42+724019.2\tablenotemark{*} & 53.0$\pm$10.5   &   \\			
 31  &    02285939+7239593 & 67.4$\pm$7.5    & \\			   
 32  &    02290968+7253475 & 21.5$\pm$1.7    &  \\			   
 33  &    02291304+7301253 & 14.8$\pm$2.5    &  \\			   
 34  &    02291769+7243426 & 64.4$\pm$4.9    & \\			   
 35  &    02291961+7302237 & 51.3$\pm$5.0    &  \\			   
 36  &    02292109+7258120 & 17.8$\pm$0.8    &  [KOS94] HA 9, T4  \\	   
 37  &    02292440+7256114 & 57.9$\pm$20.8   &  \\			   
 38  &  022932.28+725503.2\tablenotemark{*} & 72.0$\pm$10.0   &    \\			   
 39  &    02294589+7257489 & 15.5$\pm$8.0    &  \\			   
 40  &    02295744+7301240 & 101.6$\pm$17.6  &  \\			   
 41  &    02295974+7237566 & 44.4$\pm$5.4    &  \\			   
 42  &    02301621+7259542 & 18.9$\pm$4.3    &  \\			   
 43  &    02303247+7259177 & 39.4$\pm$2.6    &   IRAS 02249+7230, RNO 8, T6  \\    
 44  &    02303659+7300233 & 43.9$\pm$4.9    &   \\			
 45  &    02303676+7248328 & 85.2$\pm$13.0   &    \\			
 46  &    02303681+7259566 & 60.8$\pm$6.0    &  RNO8W,  T7  \\		
 47  &    02303717+7230370 & 41.7$\pm$4.8    &   \\			
 48  &    02303894+7236436 & 185.7$\pm$81.9  &       \\ 			
 49  &    02303911+7259572 & 29.1$\pm$1.7    &  RNO8E,  T8  \\		
 50  &    02304212+7300138 & 33.7$\pm$5.9    &    \\			     
 51  &    02304920+7239258 & 93.02$\pm$9.0   &    \\			     
 52  &    02305193+7239203 & 30.1$\pm$4.3    &   \\			     
 53  &    02305339+7243118 & 70.9$\pm$20.9   &    \\			     
 54  &    02310288+7254584 & 15.0$\pm$5.0    &    \\			     
 55  &    02310308+7237443 & 108.7$\pm$16.4  &   \\			     
 56  &    02310312+7237494 & 69.2$\pm$7.0    &   \\			     
 57  &    02310688+7243560 & 64.6$\pm$4.5    &  \\			     
 58  &    02310748+7238399 &  \nodata	       &   \\			     
 59  &    02310858+7238295 &   \nodata         &   \\			     
 60  &    02311569+7239507 & 166.1$\pm$41.5  &  \\			     
 61  &    02311975+7241146 & 36.6$\pm$5.5    &  \\			     
 62  &    02313341+7311247 & 36.14$\pm$21.7  &  \\			     
 63  &    02313994+7241575 & 33.7$\pm$6.0    &  RNO9\,B,  T9  \\		     
 64  &    02321260+7230136 &	\nodata        &       \\ 		     
 65  &    02321815+7235169 & 30.8$\pm$3.7    &   \\			     
 66  &    02322373+7239526 & 7.8$\pm$0.6     &    \\			     
 67  &    02322785+7239053 & 31.0$\pm$5.0    &   \\			     
 68  &    02322984+7236089 & 76.5$\pm$23.8   &   \\			     
 69  &    02323707+7238305 & 17.49$\pm$13.0  &   \\			     
 70  &    02323897+7239038 & 57.8$\pm$6.0    &  IRAS F02279+7225, [KOS94] HA 10, T10  \\	   
 71  &    02324009+7238545 & 147.0$\pm$14.0  &   \\			     
 72  &    02330153+7243269 & 496.0$\pm$110.0 &  IRAS 02283+7230, [KOS94] HA 11, V1180 Cas, T11  \\	    
 73  &    02330322+7243521 & 22.10$\pm$6.1   &    \\			     
 74  &    02331104+7243359 & 34.2$\pm$6.0    &     \\
 75  &    02332724+7246571 & 30.48$\pm$5.63  &      \\  		     
 \enddata  
\tablenotetext{*}{SDSS DR9 Id.}
\end{deluxetable}

\begin{deluxetable}{rccccrrc}
\tabletypesize{\scriptsize}
\tablecolumns{8}
\tablewidth{0cm}
\tablecaption{Results for the B, A, F type stars associated with L1340 \label{Table_abf}}
\tablehead{
\colhead{No.} & \colhead{2MASS Id} & \colhead{Sp. type} &  \colhead{$A_\mathrm{V}$} & \colhead{$T_\mathrm{eff}$}
 & \colhead{$L_\mathrm{star}$} & \colhead{Mass} & HRD \\
 &&& \colhead{(mag)} & \colhead{(K)} &  \colhead{(L$_{\sun}$)} & \colhead{(M$_{\sun}$)}}  
\startdata 
  I1 & 02273807+7238267  & A0  &  0.5  &    $9700^{+700}_{-500}$    &   $50.6^{+15.8}_{-12.0}$    &  2.5 &  PMS \\   
  I2 & 02280942+7237162  & F2  &  0.5  &    $6810^{+0220}_{-90}$    &    $4.4^{+1.4}_{-1.0}$	   &  1.5 & ZAMS  \\   
 I3 & 02281033+7302197  & B5  &  0.9  &   $15700^{+1000}_{-1200}$  &  $512.0^{+159.6}_{-121.7}$  &  5.0 & ZAMS \\   
 I4 & 02282291+7237539  & A5  &  1.2  &    $8080^{+190}_{-80}$      &   $13.1^{+4.1}_{-3.1}$	      &  2.0 &  ZAMS \\   
  I5 & 02290319+7259366  & A0  &  1.3  &    $9700^{+700}_{-500}$    &   $56.0^{+17.5}_{-13.3}$    &  2.5 &  PMS \\   
  I6 & 02291684+7305197  & B4  &  1.4  &   $16700^{+300}_{-1000}$   &  $649.0^{+202.3}_{-154.3}$  &  5.0 &  ZAMS \\   
  I7 & 02292000+7304545  & A3  &  0.9  &    $8550^{+290}_{-280}$    &   $20.5^{+6.4}_{-4.9}$	       &  2.0 &  ZAMS \\   
  I8 & 02292062+7304514  & A5  &  1.0  &    $8080^{+190}_{-80}$	  &   $12.1^{+3.8}_{-2.9}$	   &  2.0 &  ZAMS \\   
  I9 & 02304997+7301100  & A2  &  1.2  &    $8840^{+360}_{-80}$	  &   $68.2^{+21.3}_{-16.2}$    &  2.5 &  PMS \\   
  I10 & 02305176+7259481  & B8  &  2.7  &    $6640^{+80}_{-130}$	  &  $248.3^{+77.4}_{-59.0}$    &  3.5 &  PMS \\   
 I11 & 02314031+7241419  & F4  &  3.5  &   $12500^{+1500}_{-1800}$  &   $21.3^{+6.6}_{-5.1}$	   &  2.0 &  PMS \\  
\enddata  					     	       
\end{deluxetable}

\begin{deluxetable}{rllcc}
\tabletypesize{\scriptsize}
\tablecolumns{5}
\tablewidth{0cm}
\tablecaption{Spectral types and H$\alpha$ equivalent widths of the T~Tauri type targets of our longslit observations.  \label{Table_spres}}
\tablehead{\colhead{No.} & \colhead{2MASS Id} & \colhead{Sp. Type} & \colhead{EW(H$\alpha$)} 
&  \colhead{EW(H$\alpha$)(WFGS2)\tablenotemark{*}}}
\startdata 
  T1 &  02281661+7237328 &  G2  &  $26.4\pm0.8$    & $18.5\pm0.9$   \\
  T2 &  02281782+7238009 &  K4  &  $43.2\pm0.7$    &  $42.0\pm4.2$ \\
  T3 &  02285180+7239143 &  K4  &  $109.0\pm11.0$  & $84.0\pm4.6$   \\
  T4 &  02292109+7258120 &  G4  &  $20.5\pm1.5$    & $17.8\pm0.8$   \\
  T5 &  02293037+7311429 &  K6  &  $76.0\pm10.0$   & \nodata  \\
  T6 &  02303247+7259177 &  G7  &  $33.6\pm1.5$    & $39.4\pm2.6$  \\
  T7 &  02303681+7259566 &  K5  &  $62.6\pm2.0$    & $60.8\pm6.0$   \\
  T8 &  02303911+7259572 &  G1  &  $20.3\pm1.2$    & $29.1\pm1.7$  \\
  T9 &  02313994+7241575 &  K5  &  $16.2\pm0.6$    &  $33.7\pm6.0$  \\
  T10 &  02323897+7239038 &  K0  &  $55.9\pm2.0$\tablenotemark{a}  &  $57.8\pm6.0$  \\
  T10 &  02323897+7239038 &  K0  &  $45.0\pm1.2$\tablenotemark{b}  &  \\
  T11 &  02330153+7243269 &  K7  &  $300.0\pm15.0$  &  $496.0\pm110$   \\
  T12 &  02350799+7251034 &  K4  &  $125.5\pm12.0$\tablenotemark{a}   & $115.9\pm7.3$ \\
  T12 &  02350799+7251034 &  K4  &  $109.0\pm5.0$\tablenotemark{b}    & \nodata     \\
\enddata        				  
\tablenotetext{*}{Same as listed for these stars in Table~\ref{Table_halpha1}.}
\tablenotetext{a}{Measurred in the CAFOS spectrum.}
\tablenotetext{b}{Measured in the FAST spectrum.}
\end{deluxetable}

\begin{deluxetable}{ccccc} 
\tabletypesize{\scriptsize}
\tablecolumns{5} 
\tablewidth{0pc} 
\tablecaption{Properties of the major clumps of L1340 \label{Table_clump}}
\tablehead{ 
\colhead{Clump} & \colhead{Size} & \colhead{$\langle A_\mathrm{V} \rangle$} & \colhead{$A_\mathrm{V}$(max)} & \colhead{Mass} \\ 
 & \colhead{(pc)} & \colhead{(mag)} & \colhead{(mag)} & \colhead{(M$_{\sun}$)} }
\startdata 
A &  1.9 &  2.5 &  $> 6.7$  &  185 \\
B &  5.0 &  2.6 &  $> 6.7$  & 1325 \\ 
C &  3.4 &  2.6 &  $> 6.7$  &  620 \\
\enddata  
\end{deluxetable}

%
%
\begin{deluxetable}{rcccccccc}
\tabletypesize{\scriptsize}
\tablecolumns{9}
\tablewidth{0cm}
\tablecaption{Properties of the H$\alpha$ emission stars derived from the photometric and EW(H$\alpha$) data 
\label{Table_halpha2}}
\tablehead{\colhead{No.} & \colhead{2MASS /SDSS} & \colhead{Sp.type} & \colhead{A$_\mathrm{V}$}  &  
\colhead{T$_\mathrm{eff}$} & \colhead{L$_\mathrm{star}$}  & 
\colhead{M$_\mathrm{*}$}  & \colhead{$\log \dot{M}_\mathrm{acc}$} &  \colhead{SED Class} \\
 & & &  \colhead{(mag)} & \colhead{(K)} & \colhead{($L_{\sun}$)} & \colhead{($M_{\sun}$)} 
 }
\startdata 												    
  1  &  02263797+7304575   &  K4   &   0.7   &   $4330^{+430}_{-310}$  &   $0.35^{+0.25}_{-0.14}$    &  $1.10^{+0.20}_{-0.30}$  &    $-$8.5	 &  II\,P   \\ 
  2  &  02265909+7240166   &  M1   &   1.5   &   $3630^{+250}_{-270}$  &   $0.16^{+0.11}_{-0.06}$    &  $0.40^{+0.10}_{-0.10}$  &    $-$9.6	 &  II\,A   \\ 
  3  &  02270033+7247439   &  K3   &   1.3   &   $4550^{+370}_{-410}$  &   $0.86^{+0.61}_{-0.36}$    &  $1.25^{+0.20}_{-0.20}$  &    $-$8.2 	 &  II\,T   \\ 
  4  &  02270211+7243289   &  M0   &   2.2   &   $3770^{+170}_{-280}$  &   $0.38^{+0.27}_{-0.16}$    &  $0.55^{+0.05}_{-0.15}$  &    $-$9.1	 &  II\,A   \\ 
  5  &  02270324+7239529   &  M0   &   2.2   &   $3770^{+250}_{-270}$  &   $0.13^{+0.09}_{-0.05}$    &  $0.40^{+0.10}_{-0.10}$  &    $-$8.4 	 &  II\,A   \\ 
  6  &  02270625+7240113   &  M0   &   1.8   &   $3770^{+170}_{-280}$  &   $0.30^{+0.21}_{-0.12}$    &  $0.55^{+0.05}_{-0.15}$  &    $-$8.9 	 &  II\,A   \\  
  7  &  02273962+7250548   &  K3   &   1.2   &   $4550^{+370}_{-410}$  &   $0.99^{+0.70}_{-0.41}$    &  $1.20^{+0.20}_{-0.20}$  &    $-$8.6 	 &  II\,T   \\  
  8  &  02274044+7247536   &  K4   &   1.4   &   $4330^{+430}_{-310}$  &   $1.32^{+0.93}_{-0.54}$    &  $1.00^{+0.15}_{-0.20}$  &    $-$8.3	 &  II\,T   \\ 
  9  &  02274486+7242119   &  M0   &   1.2   &   $3770^{+170}_{-280}$  &   $0.28^{+0.20}_{-0.12}$    &  $0.55^{+0.05}_{-0.15}$  &    $-$8.8	 &  II\,P   \\  
 10  &  02275138+7309197   &  K9   &   1.2   &   $3940^{+80}_{-170}$   &   $0.12^{+0.08}_{-0.05}$    &  $0.60^{+0.10}_{-0.10}$  &    $-$10.0 	 &  III   \\
 11  &  02275976+7235561   &  K7   &   0.6   &   $3970^{+170}_{-90}$   &   $0.03^{+0.02}_{-0.01}$    &  $0.60^{+0.10}_{-0.10}$  &    $-$9.4	 & II\,P  \\
 12  &  02280135+7302439   &  K7   &   3.8   &   $3970^{+170}_{-90}$   &   $1.15^{+0.81}_{-0.48}$    &  $0.70^{+0.10}_{-0.10}$  &    $-$7.4	 &  II\,A   \\ 
 13  &  02280700+7237345   &  K6   &   2.9   &   $4020^{+210}_{-80}$   &   $0.44^{+0.31}_{-0.18}$    &  $0.80^{+0.10}_{-0.10}$  &    $-$8.4 	 &  II\,P   \\
 14  &  02281182+7236447   &  K6   &   4.6   &   $4020^{+210}_{-80}$   &   $3.28^{+2.32}_{-1.36}$    &  $0.70^{+0.10}_{-0.10}$  &    $-$7.1	 &  II\,P   \\
 15  &  02281259+7237067   &  M0   &   3.0   &   $3770^{+170}_{-280}$  &   $0.26^{+0.18}_{-0.11}$    &  $0.55^{+0.05}_{-0.15}$  &   \nodata	 &  II\,A \\
 16  &  02281661+7237328   &  G2   &   3.8   &   $5870^{+180}_{-250}$  &  $14.56^{+10.29}_{-6.03}$   &  $2.20^{+0.20}_{-0.20}$  &    $-$6.8 	  &  Flat  \\	 
 17  &  02281748+7237384   &  K4   &   4.0   &   $4330^{+430}_{-310}$  &   $1.09^{+0.77}_{-0.45}$    &  $1.10^{+0.10}_{-0.30}$  &    $-$8.0 	 &  II\,A   \\
 18  &  02281782+7238009   &  K1   &   3.8   &   $4920^{+250}_{-370}$  &   $5.27^{+3.73}_{-2.18}$    &  $2.10^{+0.40}_{-0.40}$  &    $-$7.4	 &  Flat  \\  
 19  &  02281808+7237437   &  K0   &   4.0   &   $5030^{+180}_{-270}$  &   $3.76^{+2.65}_{-1.55}$    &  $1.90^{+0.10}_{-0.20}$  &    $-$7.6 	 &  II\,P   \\
 20  &  02281818+7238069   &  K5   &   2.9   &   $4140^{+410}_{-170}$  &   $3.49^{+2.46}_{-1.44}$    &  $0.80^{+0.40}_{-0.15}$  &    $-$7.5 	 &  II\,P   \\ 
 21  &  02281877+7238091   &  K8   &   2.4   &   $3940^{+80}_{-170}$   &   $1.78^{+1.26}_{-0.74}$    &  $0.60^{+0.10}_{-0.10}$  &   \nodata	 &  II\,A    \\
 22  &  02282357+7237317   &  K8   &   3.3   &   $3940^{+80}_{-170}$   &   $1.27^{+0.89}_{-0.52}$    &  $0.60^{+0.10}_{-0.10}$  &    $-$8.0	 &  II\,A   \\
 23  &  02282383+7243450   &  M3   &   1.7   &   $3360^{+270}_{-480}$  &   $0.21^{+0.15}_{-0.09}$    &  $0.30^{+0.10}_{-0.15}$  &    $-$9.2 	 &  II\,A   \\
 24  &  02283311+7305182   &  K8   &   3.5   &   $3940^{+80}_{-170}$   &   $1.03^{+0.73}_{-0.43}$    &  $0.60^{+0.10}_{-0.10}$  &    $-$8.2 	 &  II\,A   \\ 
 25  &  02283719+7237061   &  M1   &   2.3   &   $3630^{+250}_{-270}$  &   $1.33^{+0.94}_{-0.55}$    &  $0.40^{+0.20}_{-0.10}$  &    $-$8.3 	 &  II\,A   \\
 26  &  02284780+7254422   &  M1   &   2.8   &   $3630^{+250}_{-270}$  &   $0.40^{+0.28}_{-0.16}$    &  $0.40^{+0.20}_{-0.10}$  &    $-$8.4	 &  II\,A   \\ 
 27  &  02285180+7239143   &  K4   &   1.9   &   $4330^{+430}_{-310}$  &   $1.60^{+1.13}_{-0.66}$    &  $1.10^{+0.40}_{-0.30}$  &    $-$7.6 	 &  II\,P   \\    
 28  &  02285420+7238352   &  K0   &   3.8   &   $5030^{+180}_{-270}$  &   $0.81^{+0.57}_{-0.34}$    &  $1.10^{+0.10}_{-0.10}$  &    $-$8.2	 &  II\,T   \\
 29  &  02285635+7240171   &  K6   &   2.8   &   $4020^{+210}_{-80}$   &   $1.66^{+1.17}_{-0.69}$    &  $0.70^{+0.20}_{-0.10}$  &    $-$7.3 	 &  II\,A   \\ 
 30  & 022856.42+724019.2  &\nodata & \nodata&       \nodata           &    \nodata		     &    \nodata               & \nodata	 &   \nodata	 \\
 31  &  02285939+7239593   &  K3   &   3.2   &   $4550^{+370}_{-410}$  &   $2.93^{+2.07}_{-1.21}$    &  $1.50^{+0.30}_{-0.50}$  &    $-$7.3 	 &  II\,A   \\    
 32  &  02290968+7253475   &  M0   &   1.4   &   $3770^{+170}_{-280}$  &   $0.87^{+0.62}_{-0.36}$    &  $0.55^{+0.10}_{-0.20}$  &    $-$8.3	 &  II\,T   \\    
 33  &  02291304+7301253   &  M2   &   1.6   &   $3490^{+180}_{-330}$  &   $0.15^{+0.11}_{-0.06}$    &  $0.35^{+0.05}_{-0.15}$  &    $-$9.6 	 &  II\,A   \\    
 34  &  02291769+7243426   &  K7   &   1.7   &   $3970^{+170}_{-90}$   &   $1.39^{+0.98}_{-0.58}$    &  $0.65^{+0.05}_{-0.15}$  &    $-$7.5 	 &  II\,P   \\    
 35  &  02291961+7302237   &  K5   &   6.5   &   $4140^{+410}_{-170}$  &   $1.85^{+1.30}_{-0.76}$    &  $0.80^{+0.30}_{-0.10}$  &    $-$7.6 	 &  Flat  \\	  
 36  &  02292109+7258120   &  G3   &   2.0   &   $5740^{+250}_{-230}$  &   $3.89^{+4.48}_{-2.63}$    &  $1.60^{+0.20}_{-0.20}$  &    $-$7.5 	 &  II\,T   \\     
 37  &  02292440+7256114   &  K9   &   3.0   &   $3880^{+90}_{-250}$   &   $1.72^{+1.22}_{-0.71}$    &  $0.60^{+0.05}_{-0.20}$  &    $-$7.5	 &  Flat  \\	  
 38  &  022932.32+725503.3 &  G8   &   3.0   &   $5210^{+170}_{-200}$  &	 \nodata	       &  $0.55^{+0.05}_{-0.15}$  &    $-$10.1     & \nodata   \\   
 39  &  02294589+7257489   &  M0   &   2.6   &   $3770^{+170}_{-280}$  &   $0.44^{+0.31}_{-0.18}$    &  $0.55^{+0.05}_{-0.15}$  &    $-$8.9 	 &  II\,A   \\    
 40  &  02295744+7301240   &  M0   &   3.0   &   $3770^{+170}_{-280}$  &   $0.56^{+0.40}_{-0.23}$    &  $0.65^{+0.05}_{-0.15}$  &    $-$7.4 	 &  II\,P   \\    
 41  &  02295974+7237566   &  K8   &   1.2   &   $3940^{+80}_{-170}$   &   $0.46^{+0.33}_{-0.19}$    &  $0.60^{+0.10}_{-0.20}$  &    $-$8.5 	 &  II\,T   \\    
 42  &  02301621+7259542   &  K9   &   2.5   &   $3880^{+90}_{-250}$   &   $1.25^{+0.89}_{-0.52}$    &  $1.00^{+0.10}_{-0.10}$  &    $-$8.2 	 &  II\,A   \\    
 43  &  02303247+7259177   &  G7   &   0.5   &   $5290^{+210}_{-170}$  &   $0.70^{+0.50}_{-0.29}$    &  $0.65^{+0.20}_{-0.10}$  &    $-$8.6 	 &  I	  \\	  
 44  &  02303659+7300233   &  K7   &   2.6   &   $3970^{+170}_{-90}$   &   $0.64^{+0.45}_{-0.26}$    &  $0.80^{+0.30}_{-0.10}$  &    $-$8.3 	 &  III   \\	  
 45  &  02303676+7248328   &  K5   &   4.2   &   $4140^{+410}_{-170}$  &   $1.05^{+0.74}_{-0.43}$    &  $0.70^{+0.15}_{-0.10}$  &    $-$7.7	 &  II\,P   \\    
 46  &  02303681+7259566   &  K5   &   1.2   &   $4140^{+410}_{-170}$  &   $0.84^{+0.59}_{-0.35}$    &  $0.80^{+0.40}_{-0.10}$  &    $-$8.0	 &  II\,A   \\    
 47  &  02303717+7230370   &  K6   &   2.6   &   $4020^{+210}_{-80}$   &   $1.20^{+0.85}_{-0.50}$    &  $0.65^{+0.25}_{-0.05}$  &    $-$7.9	 &  II\,A   \\    
 48  &  02303894+7236436   &  M1   &   2.0   &   $3630^{+250}_{-270}$  &   $0.56^{+0.39}_{-0.23}$    &  $0.40^{+0.20}_{-0.10}$  &    $-$7.4 	 &  III   \\	  
 49  &  02303911+7259572   &  G1   &   4.5   &   $5970^{+120}_{-230}$  &  $13.93^{+9.84}_{-5.77}$    &  $2.00^{+0.20}_{-0.20}$  &    $-$7.3 	 &  II\,P   \\    
 50  &  02304212+7300138   &  K9   &   2.0   &   $3880^{+90}_{-250}$   &   $0.67^{+0.47}_{-0.28}$    &  $0.60^{+0.10}_{-0.20}$  &    $-$8.4	 &  II\,T   \\    
 51  &  02304920+7239258   &  M1   &   2.0   &   $3630^{+250}_{-270}$  &   $0.38^{+0.27}_{-0.16}$    &  $0.40^{+0.20}_{-0.20}$  &    $-$8.0 	 &  II\,A   \\    
 52  &  02305193+7239203   &  K4   &   2.4   &   $4330^{+430}_{-310}$  &   $1.29^{+0.91}_{-0.53}$    &  $1.10^{+0.30}_{-0.30}$  &    $-$8.1 	 &  II\,A   \\    
 53  &  02305339+7243118   &  K7   &   3.0   &   $3970^{+170}_{-90}$   &   $0.99^{+0.70}_{-0.41}$    &  $0.80^{+0.15}_{-0.10}$  &    $-$7.8 	 &  II\,P   \\    
 54  &  02310288+7254584   &  K9   &   2.5   &   $3880^{+90}_{-250}$   &   $0.46^{+0.32}_{-0.19}$    &  $0.60^{+0.10}_{-0.20}$  &    $-$9.0 	 &  II\,A   \\    
 55  &  02310308+7237443   &  K3   &   3.0   &   $4550^{+370}_{-410}$  &   $3.21^{+2.27}_{-1.33}$    &  $1.50^{+0.50}_{-0.50}$  &    $-$7.0	 &  II\,A   \\   
 56  &  02310312+7237494   &  K5   &   3.0   &   $4140^{+410}_{-170}$   &   $0.96^{+0.68}_{-0.40}$   &  $0.60^{+0.10}_{-0.10}$  &    $-$7.8	 &  II\,P   \\    
 57  &  02310688+7243560   &  K5   &   2.6   &   $4140^{+410}_{-170}$  &   $1.99^{+1.40}_{-0.82}$    &  $0.80^{+0.30}_{-0.20}$  &    $-$7.4 	 &  II\,A   \\    
 58  &  02310748+7238399   &  K9   &   2.8   &   $3880^{+90}_{-250}$   &   $0.42^{+0.29}_{-0.17}$    &  $0.60^{+0.10}_{-0.15}$  &    \nodata	 &  II\,A  \\	  
 59  &  02310858+7238295   &  M0   &   2.7   &   $3770^{+170}_{-280}$  &   $0.74^{+0.53}_{-0.31}$    &  $0.55^{+0.05}_{-0.15}$  &    \nodata	 &  II\,T  \\	  
 60  &  02311569+7239507   &  M2   &   2.8   &   $3490^{+180}_{-330}$  &   $0.22^{+0.15}_{-0.09}$    &  $0.35^{+0.05}_{-0.10}$  &    $-$8.2 	 &  II\,A   \\    
 61  &  02311975+7241146   &  K8   &   2.6   &   $3940^{+80}_{-170}$   &   $0.84^{+0.59}_{-0.35}$    &  $0.60^{+0.04}_{-0.10}$  &    $-$8.2 	 &  II\,A   \\    
 62  &  02313341+7311247   &  K3   &   2.6   &   $4550^{+370}_{-410}$  &   $1.06^{+0.75}_{-0.44}$    &  $1.30^{+0.15}_{-0.15}$  &    $-$8.3	 &  II\,A   \\    
 63  &  02313994+7241575   &  K5   &   3.5   &   $4140^{+410}_{-170}$  &   $1.24^{+0.88}_{-0.51}$    &  $0.80^{+0.40}_{-0.20}$  &    $-$8.0 	 &  II\,P   \\    
 64  &  02321260+7230136   &  K2   &   3.5   &   $4760^{+270}_{-430}$  &   $0.96^{+0.68}_{-0.40}$    &  $1.20^{+0.30}_{-0.20}$  &    \nodata	 &  II\,P    \\   
 65  &  02321815+7235169   &  K0   &   3.8   &   $5030^{+180}_{-330}$  &   $1.57^{+1.11}_{-0.65}$    &  $1.50^{+0.30}_{-0.30}$  &    $-$8.2 	 &  II\,A   \\    
 66  &  02322373+7239526   &  K4   &   2.0   &   $4330^{+370}_{-310}$  &   $2.89^{+4.00}_{-2.34}$    &  $1.50^{+0.30}_{-0.30}$  &    $-$7.6 	 &   II\,P   \\ 
 67  &  02322785+7239053   &  M3   &   2.0   &   $3360^{+270}_{-480}$  &   $0.25^{+0.18}_{-0.11}$    &  $0.30^{+0.10}_{-0.20}$  &    \nodata	 &   III    \\
 68  &  02322984+7236089   &  K7   &   2.8   &   $3970^{+170}_{-90}$   &   $0.98^{+0.69}_{-0.41}$    &  $0.65^{+0.15}_{-0.05}$  &    $-$7.7	 &   II\,A   \\
 69  &  02323707+7238305   &  K8   &   3.0   &   $3940^{+80}_{-170}$   &   $1.52^{+1.07}_{-0.63}$    &  $0.60^{+0.05}_{-0.10}$  &    $-$8.0 	 &   II\,A   \\
 70  &  02323897+7239038   &  K0   &   3.4   &   $5030^{+180}_{-270}$  &   $6.77^{+4.78}_{-2.80}$    &  $2.30^{+0.20}_{-0.20}$  &    $-$6.9	 &   II\,P   \\   
 71  &  02324009+7238545   &  M2   &   1.6   &   $3490^{+180}_{-330}$  &   $0.21^{+0.15}_{-0.09}$    &  $0.35^{+0.10}_{-0.12}$  &    $-$8.4 	 &   III   \\
 72  &  02330153+7243269   &  K7   &   4.6   &   $3970^{+170}_{-90}$   &   $3.19^{+2.25}_{-1.32}$    &  $0.65^{+0.15}_{-0.05}$  &    $-$5.8 	 &   Flat  \\	
 73  &  02330322+7243521   &  M2   &   4.2   &   $3490^{+180}_{-330}$  &   $1.57^{+1.11}_{-0.65}$    &  $0.35^{+0.10}_{-0.10}$  &    $-$8.3 	 &   II\,A   \\
 74  &  02331104+7243359   &  M2   &   2.8   &   $3490^{+180}_{-330}$  &   $0.39^{+0.27}_{-0.16}$    &  $0.35^{+0.10}_{-0.10}$  &    $-$8.7	 &   II\,A   \\
 75  &  02332724+7246571   &  K2   &   4.8   &   $4760^{+270}_{-430}$  &   $1.31^{+0.93}_{-0.54}$    &  $1.40^{+0.30}_{-0.20}$  &    $-$8.3	 &   II\,P   \\
 76\tablenotemark{a} & 02293037+7311429 &  K6 & 3.8 &  $4020^{+210}_{-80}$   &   $3.27^{+2.31}_{-1.35}$    &  $0.70^{+0.20}_{-0.05}$  &    $-$6.9      &  II\,P  \\     
 77\tablenotemark{b} & 02350799+7251034 &  K4 & 1.8 &  $4330^{+430}_{-310}$  &   $1.56^{+1.10}_{-0.65}$    &  $1.10^{+0.30}_{-0.30}$  &    $-$7.4      &   II\,P  \\
\enddata  				         		    	                		   
\tablenotetext{a}{KOS94~HA\,4, outside of the field of view of the WFGS2 observations}												       
\tablenotetext{b}{ KOS94~HA\,13, outside of the field of view of the WFGS2 observations }
\end{deluxetable}

\begin{deluxetable}{lcccccccc} 
\tabletypesize{\scriptsize}
\tablecolumns{8} 
\tablewidth{0pc} 
\tablecaption{Distribution of SED slopes of the candidate PMS stars \label{Table_sedcomp}}
\tablehead{ 
\colhead{SED Class} & \colhead{$\langle A_\mathrm{V} \rangle$ } & \colhead{$\langle EW(\ha) \rangle$} & \colhead{$\langle T_\mathrm{eff} \rangle$} & \colhead{$\langle M_{star} \rangle$} & \colhead{$\langle L_{star} \rangle$} & \colhead{$\langle \dot{M}_\mathrm{acc} \rangle$} & \colhead{N} \\
 & \colhead{(mag)} & \colhead{(\AA)} & \colhead{(K)} & \colhead{(M$_{\sun}$)} & \colhead{(L$_{\sun}$)} & \colhead{($10^{-9}$\,M$_{\sun}$\,yr$^{-1}$)} }
\startdata  
I + Flat      & 3.80  & 135.0 & 4680 & 1.24  & 4.90 & 285.3 & 6 \\
\phn \phn II P   & 2.80  & 75.0  & 4340 & 1.00  & 2.28 & 31.9  & 23 \\
\phn \phn II T   & 1.84  & 33.4  & 4380 & 0.96  & 1.54 & 8.3   & 9  \\
\phn \phn II A   & 2.68  & 49.4  & 3910 & 0.68  & 1.00 & 15.5  & 32 \\	
II (P+T+A)    & 2.63  & 57.1  & 4140 & 0.83  & 1.98 & 21.7  & 64 \\
III           & 1.88  & 14.0  & 3670 & 0.46  & 0.47 & 11.4  &  5 \\
\enddata  
\end{deluxetable} 

\begin{deluxetable}{lcccc} 
\tabletypesize{\scriptsize}
\tablecolumns{5} 
\tablewidth{0pc} 
\tablecaption{Comparison of Class~II YSOs in L1340 and Taurus \label{Table_taucomp}} 
\tablehead{
 & \colhead{L1340 H$\alpha$} & \colhead{L1340 Spitzer\tablenotemark{a}}  & \colhead{Taurus Spitzer} & \colhead{Taurus ref.}} 
\startdata  
Stars earlier than K6            & 26      & 37      & 23	 &  1 \\
K6 $\le Sp. type <$ M3            & 38      & 71      & 90	 &  1 \\
M3 $\le Sp. type <$ M6            &  2      & 44      & 44	 &  1 \\
Mean $K_\mathrm{s}-[5.8]$         & 1.52    & 1.29    & 1.38	 &  1 \\
Mean $K_\mathrm{s}-[8.0]$         & 2.24    & 1.94    & 2.20	 &  1 \\
Mean $K_\mathrm{s}-[24 ]$         & 4.96    & 4.78    & 5.16	 &  1 \\
Mean $\log \dot{M}_{acc}$         & $-8.1$   & \nodata   & $-7.6$   &  2 \\
\phn \phn Primordial disks          & $-7.6$   & \nodata   & $-7.6$   &  2 \\
\phn \phn Transitional disks        & $-8.1$   & \nodata   & $-8.5$   &  2 \\
\enddata  
\tablenotetext{a}{Paper III}
\tablerefs{(1) \citet{Luhman2010}; (2) \citet{Najita2007}}
\end{deluxetable} 

\clearpage

\begin{deluxetable}{ccccccccc}
\tabletypesize{\scriptsize}
\tablecolumns{9}
\tablewidth{0cm}
\rotate
\tablecaption{Table A1. \textit{UBVR$_\mathrm{C}$I$_\mathrm{C}$JHK$_s$} magnitudes of the optically selected candidate YSOs of L1340 \label{Tab_phot1}}
\tablehead{
\colhead{N} & \colhead{U ~~dU} &  \colhead{B ~~dB} & \colhead{V ~~dV} & \colhead{R$_\mathrm{C}$ ~~dR$_\mathrm{C}$} & 
\colhead{I$_\mathrm{C}$ ~~dI$_\mathrm{C}$} & \colhead{J ~~dJ}  & \colhead{H ~~dH} & \colhead{K$_s$ ~~dK$_s$}}  
\startdata 
\cutinhead{Intermediate-mass stars}
  1   &     12.815   0.012 &  11.222 0.030 & 10.906 0.030 & 10.886 0.030 & 10.769  0.030   & 10.606  0.025  & 10.548  0.031  & 10.553  0.019 \\
  2   &     14.266   0.009 &  13.770 0.053 & 13.450 0.030 & 13.165 0.030 & 12.812  0.025   & 12.235  0.025  & 11.997  0.031  & 11.926  0.023 \\ 
  3   &     10.332   0.005 &  10.090 0.030 &  9.977 0.025 &  9.864 0.020 &  9.668  0.020   &  9.565  0.026  &  9.534  0.032  &  9.496  0.021 \\ 
  4   &     14.392   0.019 &  13.230 0.053 & 12.916 0.042 & 12.562 0.034 & 12.248  0.034   & 11.698  0.025  & 11.547  0.029  & 11.496  0.023 \\ 
  5   &     13.318   0.008 &  11.838 0.030 & 11.562 0.030 & 11.300 0.030 & 11.048  0.020   & 10.629  0.027  & 10.470  0.030  & 10.386  0.021 \\ 
  6   &     12.743   0.018 &  10.733 0.042 & 10.388 0.042 & 10.100 0.034 & 10.018  0.034   &  9.810  0.027  &  9.820  0.030  &  9.828  0.019 \\ 
  7   &     13.363   0.018 &  12.442 0.020 & 12.086 0.020 & 11.847 0.020 & 11.669  0.030   & 11.334  0.065  & 11.155  0.058  & 11.081  0.054 \\ 
  8   &     14.973   0.033 &  13.190 0.020 & 12.755 0.020 & 12.450 0.020 & 12.222  0.020   & 11.771  0.030  & 11.702  0.032  & 11.576  0.023 \\ 
  9   &     12.757   0.007 &  11.410 0.053 & 11.164 0.042 & 10.918 0.031 & 10.629  0.031   & 10.151  0.025  &  9.915  0.030  &  9.860  0.021 \\ 
 10   &     13.190   0.025 &  12.502 0.022 & 11.823 0.036 & 11.300 0.030 & 10.879  0.034   & 10.188  0.025  &  9.765  0.030  &  9.698  0.021 \\ 
 11   &     16.083   0.010 &  15.544 0.030 & 14.515 0.040 & 13.646 0.030 & 12.909  0.030   & 11.499  0.025  & 10.582  0.029  &  9.831  0.018  \\
\cutinhead{H$\alpha$ emission stars} 
   1  &    16.810   0.014  &   16.509	0.006  &   15.648   0.006  &   15.084	0.007  &   14.316   0.007  &  14.211  0.031  &  13.403 0.037  &   13.034 0.028   \\ 
   2  &    22.784   1.007  &   21.276	0.030  &   19.646   0.030  &   18.415	0.018  &   16.796   0.014  &  15.167  0.044  &  14.428 0.046  &   14.152 0.052   \\ 
   3  &    19.056   0.045  &   17.949	0.007  &   16.489   0.007  &   15.598	0.007  &   14.783   0.008  &  13.578  0.026  &  12.875 0.028  &   12.630 0.022   \\ 
   4  &    23.127   1.610  &   20.862	0.021  &   19.156   0.021  &   17.870	0.014  &   16.204   0.011  &  14.581  0.034  &  13.661 0.041  &   13.309 0.033   \\ 
   5  &    22.645   0.283  &   22.006	0.046  &   20.235   0.046  &   18.982	0.026  &   17.453   0.021  &  15.645  0.054  &  14.718 0.062  &   14.328 0.069   \\ 
\enddata
\tablecomments{Table A1 is published in its entirety in the
electronic edition of the {\it Astrophysical Journal}.  A portion is
shown here for guidance regarding its form and content.}
\end{deluxetable}		        		   		        		        			       
				        		  		      			     
\clearpage

\begin{deluxetable}{ccccccccccc}
\tabletypesize{\scriptsize}
\tablecolumns{11}
\tablewidth{0cm}
\rotate
\tablecaption{Table A2. \textit{Spitzer\/} and \textit{AllWISE\/} magnitudes of the optically selected candidate YSOs of L1340 \label{Tab_phot2}}
\tablehead{
\colhead{N} & \colhead{[3.6] ~d[3.6] } & \colhead{[4.5] ~d[4.5]} & \colhead{[5.8] ~d[5.8]} & \colhead{[8.0] ~d[8.0]} & \colhead{[24] ~d[24]} & \colhead{[70] ~d[70]} & 
 \colhead{W1 ~dW1} & \colhead{W2 ~dW2} & \colhead{W3 ~dW3} &  \colhead{W4 ~dW4}}  
\startdata 
\cutinhead{Intermediate-mass stars}
  1  &    10.600  0.034  &  10.624  0.034  & 10.662  0.033  & 10.697  0.037  &   \nodata	& \nodata 	  & 10.378  0.023   &  10.393 0.020   &  8.903 0.036  &  7.208 0.101  \\
  2  &    11.944  0.040  &  11.912  0.034  & 11.990  0.059  & 11.976  0.032  &   \nodata	& \nodata 	  & 11.751  0.023   &  11.720 0.021   & $\geqq$11.750 &  $\geqq$9.228	\\
  3  &     9.506  0.034  &   9.525  0.040  &  9.559  0.036  &  9.606  0.038  &   \nodata	&\nodata  	  &  9.431  0.023   &	9.451 0.020   &  9.574 0.051  &  $\geqq$8.933  \\
  4  &    11.493  0.034  &  11.487  0.035  & 11.503  0.037  & 11.525  0.037  &   \nodata	& \nodata 	  & 11.436  0.023   &  11.394 0.022   & 11.041 0.131  &  $\geqq$8.838 \\
  5  &    10.358  0.033  &  10.352  0.033  & 10.387  0.034  & 10.455  0.038  &   \nodata	& \nodata 	  & 10.337  0.023   &  10.299 0.021   &  9.519 0.128  &  8.740 0.362  \\
  6  &     9.834  0.034  &   9.830  0.033  &  9.832  0.035  &  9.817  0.034  &  8.984  0.082	& \nodata 	  &  9.816  0.022   &	9.834 0.020   &  9.769 0.058  &  $\geqq$8.668 \\
  7  &    11.063  0.034  &  11.061  0.036  & 11.073  0.036  & 10.985  0.051  &   \nodata	& \nodata 	  & 10.451  0.022   &  10.477 0.020   &  9.842 0.055  &  8.056 0.179  \\
  8  &    11.536  0.045  &  11.571  0.047  & 11.489  0.045  & 11.354  0.052  &   \nodata	& \nodata 	  & \nodata	    &  \nodata        &  \nodata      & \nodata  \\
  9  &     9.814  0.035  &   9.798  0.033  &  9.804  0.037  &  9.821  0.036  &   \nodata	& \nodata 	  &  9.794  0.022   &	9.794 0.020   &  9.162 0.060  &  $\geqq$8.499\\
 10  &     9.729  0.001  &   9.722  0.001  &  9.686  0.002  &  9.634  0.000  &   \nodata	& \nodata 	  &  9.606  0.023   &	9.662 0.020   &  9.482 0.038  &  8.279 0.325  \\
 11  &     8.496  0.053  &   8.084  0.033  &  7.764  0.033  &  7.316  0.035  &  5.809  0.044    &  8.608  0.021   &   7.998 0.017   &   6.921 0.014   &  5.670 0.033  &    \nodata  \\
\cutinhead{H$\alpha$ emission stars} 
   1  &     12.266  0.002   &  11.641  0.002  &  11.059  0.004  &  10.095  0.003  &   7.421  0.016  & 3.331 0.096   &  12.346  0.023  &   11.516   0.022  &\phn9.239 0.033  &  7.224  0.128   \\ 
   2  &     13.659  0.003   &  13.358  0.005  &  13.091  0.011  &  12.141  0.011  &   9.422  0.180  &	 \nodata    &  13.734  0.024  &   13.331   0.026  &  11.527  0.173  &  $\geqq$8.498   \\ 
   3  &     12.596  0.002   &  12.428  0.003  &  12.160  0.006  &  11.193  0.005  &   7.260  0.015  &	 \nodata    &  12.553  0.024  &   12.404   0.023  &  10.133  0.058  &  7.421   0.114   \\ 
   4  &     12.836  0.002   &  12.615  0.004  &  12.606  0.008  &  12.053  0.013  &   8.661  0.054  &	 \nodata    &  13.041  0.025  &   12.719   0.026  &  11.724  0.199  &  $\geqq$8.241    \\ 
   5  &     13.449  0.003   &  12.960  0.004  &  12.902  0.010  &  11.827  0.009  &   8.780  0.088  &	\nodata     &  13.582  0.025  &   12.950   0.025  &  11.095  0.116  &  $\geqq$8.290    \\ 
\enddata
\tablecomments{Table A2 is published in its entirety in the
electronic edition of the {\it Astrophysical Journal}.  A portion is
shown here for guidance regarding its form and content.}
\end{deluxetable}

\end{document}